\documentstyle[12pt,epsf,psfig,eqsection,cite]{article}

\def\thefootnote{\fnsymbol{footnote}}

\newcommand{\eq}{\begin{equation}}
\newcommand{\en}{\end{equation}}
\newcommand{\eqa}{\begin{eqnarray}}
\newcommand{\ena}{\end{eqnarray}}

\def\spose#1{\hbox to 0pt{#1\hss}}
\def\ltapprox{\mathrel{\spose{\lower 3pt\hbox{$\mathchar"218$}}
 \raise 2.0pt\hbox{$\mathchar"13C$}}}
\def\gtapprox{\mathrel{\spose{\lower 3pt\hbox{$\mathchar"218$}}
 \raise 2.0pt\hbox{$\mathchar"13E$}}}

\def\case#1#2{{\textstyle\frac{#1}{#2}}}

\def\bphi{\mbox{\boldmath $\phi$}}
\newcommand{\be}{\begin{equation}}
\newcommand{\ee}{\end{equation}}
\newcommand{\reff}[1]{(\ref{#1})}

\begin{document}
\begin{titlepage}
\vskip0.5cm
\begin{flushright}
\end{flushright}
\vskip0.5cm
\begin{center}
{\Large\bf Large-$n$ Critical Behavior of $O(n)\times O(m)$ Spin Models}
\vskip 0.3cm
{\Large\bf }
\end{center}
\vskip 1.3cm
\centerline{
Andrea Pelissetto,$^a$ $\quad$ Paolo Rossi,$^b$ $\quad$ 
Ettore Vicari$^b$}

 \vskip 0.4cm
 \centerline{\sl  $^a$ Dipartimento di Fisica dell'Universit\`a di Roma I
 and I.N.F.N., I-00185 Roma, Italy}
 \centerline{\sl  $^b$ Dipartimento di Fisica dell'Universit\`a di Pisa 
 and I.N.F.N., I-56127 Pisa, Italy}
 \medskip
 \centerline{\tt E-mail: Pelissetto@roma1.infn.it,
                         Rossi@df.unipi.it, Vicari@df.unipi.it}
 \vskip 1.cm

\begin{abstract}
We consider the Landau-Ginzburg-Wilson Hamiltonian with 
$O(n)\times O(m)$ symmetry and compute the critical exponents at 
all fixed points to $O(n^{-2})$ and to $O(\epsilon^3)$ in a 
$\epsilon=4-d$ expansion. We also consider the corresponding 
non-linear $\sigma$-model and determine the fixed points and 
the critical exponents to $O(\tilde{\epsilon}^2)$ in the 
$\tilde{\epsilon}=d-2$ expansion. Using these results, we draw quite 
general conclusions on the fixed-point structure of models with
$O(n)\times O(m)$ symmetry for $n$ large and all $2\le d\le 4$.
\end{abstract}

\vskip 1.cm

PACS Numbers: 05.70.Jk, 64.60.Fr, 75.10.Hk, 11.10.Kk, 11.15.Pg

\vskip 1.cm

Keywords: 

Critical phenomena, frustrated  models, O($M$)$\times$ O($N$)-symmetric models, 

Field theory,  $1/N$ expansion, $\epsilon$-expansion.

\end{titlepage}

\setcounter{footnote}{0}
\def\thefootnote{\arabic{footnote}}

\section{Introduction.}
\label{seceintro}

The critical behavior of frustrated $XY$ and Heisenberg spin systems 
with noncollinear order has been the
 subject of many recent theoretical studies, where the standard tools of 
renormalization-group (RG) theory have been applied to field theories which were conjectured to be 
appropriate for the description of the systems under investigation 
(see, e.g., Refs. \cite{KawaR,PV-01} for reviews on this issue).

The critical behavior of these systems is rather controversial. Indeed,
while experimentally there is good evidence of a second-order phase 
transition\footnote{The experimental results are reviewed in, 
e.g., Refs. \cite{KawaR,CP-97}. Essentially, experiments with hexagonal
peroskvites find a clear second-order phase transition except
for CsCuCl${}_3$. The results for helimagnetic rare earths are instead
less clear. We also mention Ref. \cite{PKVMW-00} 
where it is shown experimentally that chiral order and spin order occur 
simultaneously, thereby supporting Kawamura's \cite{Kawa-gen,Kawa2} conjecture 
that chiral transitions are different from the standard $O(n)$ transitions.}
belonging to a new (chiral) universality class, 
theoretically the issue is still debated.
On one side, field-theoretical studies based in approximate solutions of the 
RG equations (ERG) do not find any stable fixed point and favor a first-order 
phase transition \cite{Tiss1,Tiss2,Tiss3}. 
On the other hand, perturbative field 
theory  gives the opposite answer: A stable fixed point is identified 
with exponents in agreement with the experiments \cite{PRV-00}.
Monte Carlo simulations 
\cite{Kawamura-92,MC2,MC} do not help clarifying the issue. While 
simulations of the antiferromagnetic $O(n)$ model on a 
stacked triangular lattice find a second-order phase transition with 
exponents reasonably near to the experimental ones, modified spin 
systems which supposedly belong to the same universality class
apparently favor a first-order transition \cite{MC}.\footnote{
Note that not all modified models show a first-order phase transitions.
Some of them have a behavior that is compatible with a second-order 
phase transition. However, the measured exponents do not satisfy the condition
$\eta \ge 0$, which must be satisfied in unitary
(reflection-positive) models as these are, so that, 
the measured exponents can only be effective ones.
This is interpreted as a signal of a first-order phase transition.
However, there is also the possibility that the results are strongly
biased by corrections to scaling induced by the constraint.} 
Note that the existence of a new chiral universality class does not 
exclude the possibility that some systems undergo a first-order transition.
Indeed, they may lie outside the attraction domain of the stable fixed point
and thus belong to runaway RG trajectories. In this case,
a first-order transition is expected.

The field-theoretical studies have been focusing either on the so-called 
Landau-Ginzburg-Wilson (LGW) Hamiltonian with $O(n) \times O(2)$ symmetry 
or on the corresponding
nonlinear sigma (NL$\sigma$) model.  In this paper we will study a 
generalization of these theories, by considering general 
$O(n) \times O(m)$ Hamiltonians and we will try to understand the nature 
of the fixed points of the theory. In particular, we will relate 
the LGW and the NL$\sigma$ descriptions showing explicitly that 
the stable fixed points of the two models are exactly the same, 
as conjectured in Refs. \cite{Kawa3,Aza0,Aza}, 
for values of $m$ and $n$ consistent
with the existence of a second-order phase transition.
Moreover, we will clarify the nature of the unstable fixed
points. However, this analysis will only be valid in the large-$n$ 
region, where, by using the large-$n$ expansion, we will be able to identify 
nonperturbatively all fixed points of the different Hamiltonians.

Since the large-$n$ expansion plays a major role in our discussion, our results
will only be valid for $n > \overline{n}(m,d)$, i.e. in the 
region of large-$n$ analyticity. At least near four dimensions, 
such a function is conjectured to be 
identified with the line $n^+(m,d)$ on which the LGW chiral and antichiral 
fixed points merge. The function $n^+(m,d)$ has been the object of extensive 
studies that tried to understand whether, in the physical case
$d=3$ and $m=2$, $\overline{n}(m,d)$ was smaller or larger than three. 
In this case, studies using various approaches gave 
$\overline{n}(2,3) \approx 4$ (Refs. \cite{AS-94,ASV-95,Tiss2}), 
5 (Ref. \cite{Tiss3}), $\approx 6$ (Ref. \cite{PRV-00}). Here, we will 
provide another determination, together with generalizations 
for other values of $m$, that substantially confirms previous 
findings, i.e. $\overline{n}(2,3)\approx 5$.  Since the results that we will present 
are essentially  adiabatic moving  from 
the large-$n$ and small $\epsilon\equiv 4-d$ region, 
they are not expected to provide the (essentially nonperturbative) features of 
the models in the region below $\bar n(m,d)$.
Therefore, the fact that $\overline{n}(2,3) > 3$
does not necessarily imply an inconsistency
with the field-theoretical results of
Ref. \cite{PRV-00}, where a rather robust evidence for stable chiral
fixed points was found for $O(n) \times O(2)$ models with $n=2$, 3 in fixed 
dimension $d=3$.  Such fixed points are not analytically connected with the 
large-$n$ and small-$\epsilon$ criticalities discussed above.

We will also show that for $m> 2$ the identification 
$\overline{n}(m,d) = n^+(m,d)$ may not be correct for $d$ near two
dimensions. Indeed, in this case a new critical line appears 
which corresponds to the merging of the chiral fixed point with the
NL$\sigma$ antichiral fixed point. 

In order to obtain quantitative predictions for all $m$ and $n$, 
we have extended the $\epsilon \equiv 4 - d$ expansion of the LGW theory 
to order $\epsilon^3$ and the $\tilde{\epsilon} \equiv d-2$ expansion 
of the NL$\sigma$ model to order $\tilde{\epsilon}^2$ for all $n$ and $m$. 
Also, we present $O(1/n^2)$ results for the LGW theory.

The paper is organized as follows.
In Sec. \ref{sec-models} we define the general class of models 
with $O(n)\times O(m)$ symmetry that will be considered in the paper
and find a general representation that is the starting point of the 
large-$n$ expansion.
In Sec. \ref{sec-epsilon}
we compute the $O(\epsilon^3)$ contributions to the critical exponents 
$\eta$ and 
$\nu^{-1}$ within the $\epsilon=4-d$ expansion of the LGW Hamiltonian.
In Sec. \ref{sec-largen} 
we analyze in detail the $1/n$ expansion of the LGW Hamiltonian with 
$O(n) \times O(m)$ symmetry 
to O($1/n^2$), thereby extending the results of Ref. \cite{Kawa3}. 
Interestingly enough, we can determine the large-$n$ expansion of the 
exponents at all fixed points and show explicitly their different physical 
nature: at the stable fixed point both tensor and scalar excitations propagate,
while at each unstable fixed point one of the degrees of freedom is 
suppressed. At the Heisenberg fixed point there are only scalar excitations,
while at the antichiral one, there are only tensor excitations.
In Sec. \ref{sec-vector} we discuss the $1/n$ expansion of a more general
theory in which the coupling to a (gauge) 
vector field is included, extending the results  of Ref. \cite{Hika}.
In Sec. \ref{sec-sigma} 
we extend to arbitrary values of $m$ and to 
$O(\tilde\epsilon^2)$ the $\tilde\epsilon$-expansion 
of NL$\sigma$ models, evaluating the unstable fixed point and the 
coalescence value of $n$ under which the two fixed points actually disappear. 
We also identify the ``gauge'' criticality of the models. 
In Sec. \ref{sec-conclusions} we draw some general conclusions and present 
a new determination of the function $\bar{n}(m,d)$.

\section{Models.}
\label{sec-models}

We will consider a non-Abelian gauge model coupled to a scalar field with 
gauge symmetry $O(m)$ and global symmetry $O(n)$. In particular, we consider 
a set of $m$ $n$-dimensional vectors $\bphi_\alpha = \{\phi_{\alpha a}\}$, 
$\alpha = 1,\ldots, m$, $a = 1, \ldots, n$, a vector field 
$A^{\alpha\beta}_\mu$ antisymmetric in $\alpha$ and $\beta$, and the 
Hamiltonian density
\begin{eqnarray}
{\cal H} = &&{1 \over 2} \sum_{\alpha} (\partial_{\mu} 
{\bphi}_{\alpha}+g_0 A_{\mu}^{\alpha\beta}{\bphi}_{\beta})^2
 + {1 \over 2} r_0
\sum_{\alpha}  {\bphi}_{\alpha}^2 + 
   {1 \over 4!} u_0 \left(\sum_{\alpha}  {\bphi}_{\alpha}^2\right)^2 
\nonumber \\
&&+ {1 \over 4!} v_0 \sum_{\alpha \beta} 
  \left[ ({\bphi}_{\alpha}\cdot {\bphi}_{\beta})^2 -
          {\bphi}_{\alpha}^2 {\bphi}_{\beta}^2\right]
  +{1 \over 4}F_{\mu\nu}^2 + 
  {t_0\over 2} \sum_{\alpha\beta} A_\mu^{\alpha\beta} A_\mu^{\alpha\beta} ,
\label{Hgeneral}
\end{eqnarray}
where $F_{\mu\nu}$ is the non-Abelian field strength associated with the 
fields $A_{\mu}^{\alpha\beta}$. This Hamiltonian is gauge-invariant 
(with local $O(m)$ invariance) for $t_0 = 0 $, and in this case it has 
already been studied\footnote{Hikami's couplings \cite{Hika},
labelled by the subscript $H$, are related to ours by the correspondence:
$\rho_H = {(u_0-v_0)/3}$, $v_H = {v_0/6}$,
$\lambda_H \equiv \rho_H + 2 v_H = { u_0/3}$.}
in Ref. \cite{Hika}. For $t_0 = 0$ and 
$A_\mu^{\alpha\beta} = 0$  we obtain a generic LGW Hamiltonian density 
with global $O(m)\times O(n)$ invariance: 
\begin{eqnarray}
{\cal H} = &&{1 \over 2} \sum_{\alpha} (\partial_{\mu} 
{\bphi}_{\alpha})^2
 + {1 \over 2} r_0
\sum_{\alpha}  {\bphi}_{\alpha}^2 + 
   {1 \over 4!} u_0 \left(\sum_{\alpha}  {\bphi}_{\alpha}^2\right)^2 
\nonumber \\
&&+ {1 \over 4!} v_0 \sum_{\alpha \beta} 
  \left[ ({\bphi}_{\alpha}\cdot {\bphi}_{\beta})^2 -
          {\bphi}_{\alpha}^2 {\bphi}_{\beta}^2\right].
\label{H-LGW}
\end{eqnarray}
Assuming $n>m$,  stability requires $u_0>0$ and 
$w_0 \equiv u_0 + (1 - {\cal N}) v_0/{\cal N} > 0$, where 
${\cal N} = {\rm min}\ (m,n)$. 

Other particular cases
of the Hamiltonian \reff{Hgeneral} are interesting. 
If we set $u_0 = v_0$ and $r_0 = - v_0 \eta_1/6$
and take the limit $v_0\to+\infty$ keeping $\eta_1$ fixed, we obtain an 
$O(n)\times O(m)$ $\sigma$-model coupled to an $O(m)$ vector field. 
The Hamiltonian density is given by 
\eq
{\cal H} = {1 \over 2} \sum_{\alpha} (\partial_{\mu} 
{\bphi}_{\alpha}+g_0 A_{\mu}^{\alpha\beta}{\bphi}_{\beta})^2
+{1 \over 4}F_{\mu\nu}^2 + 
{t_0\over 2} \sum_{\alpha\beta} A_\mu^{\alpha\beta} A_\mu^{\alpha\beta} ,
\label{sigma-model-gauged}
\en
where the fields $\bphi$ satisfy the constraint
\begin{equation}
\bphi_\alpha \cdot \bphi_\beta = \eta_1 
   \delta_{\alpha\beta}.
\label{sec2:constraint}
\end{equation}
This limit is well defined only if $n\ge m$, otherwise the 
constraint \reff{sec2:constraint} cannot be satisfied.
In the absence of the kinetic term for the vector field, the Hamiltonian
density depends quadratically on the vector field that can then
be eliminated by integration. We obtain a new 
Hamiltonian of the form
\begin{eqnarray}
{\cal H} = {1 \over 2} \sum_{\alpha} (\partial_{\mu} {\bphi}_{\alpha})^2  - 
   {g_0^2\over 8 (g_0^2\eta_1 + t_0) }
   \sum_{\alpha \beta} 
   (\bphi_\alpha \partial_\mu\bphi_\beta - 
    \bphi_\beta \partial_\mu\bphi_\alpha)^2.
\end{eqnarray}
This is the $\sigma$-model studied in Refs. \cite{Kawa3,Aza0,Aza}. 
In order to recover the notations\footnote{Notice 
that our $\eta_i$ are consistent with the couplings employed
in Sec. 4 of Ref. \cite{Aza}, but they are twice as big as the couplings 
defined in Appendix B of the same paper, due to a 
slight inconsistency in the notation adopted by these authors.
The couplings $\eta_i$ are related to those of Ref. \cite{Kawa3} by
$\eta_1 \equiv {1/T}+ {1/T^\prime}$,
$\eta_2 \equiv {2/T}$.
For easy comparison, the reader should keep in mind that, 
due to the constraints,
$
  {\bf e}_\alpha  \cdot \partial_\mu{\bf e}_\beta = 
- {\bf e}_\beta  \cdot \partial_\mu{\bf e}_\alpha
$.}
of Ref. \cite{Aza}, we set 
$\bphi_\alpha = \sqrt{\eta_1} {\bf e}_\alpha$ and 
$\eta_2 \equiv 2 t_0 \eta_1 /(g_0^2 \eta_1+ t_0)$. Then
\eq
\tilde H = 
  {1 \over 2}\eta_1 \sum_{\alpha=1}^m 
  \partial_\mu{\bf e}_\alpha \cdot \partial_\mu{\bf e}_\alpha +
 \left({1 \over 2}\eta_2-\eta_1\right) 
 \sum_{\alpha>\beta}^m ({\bf e}_\alpha \cdot\partial_\mu{\bf e}_\beta)^2,
\label{Hamiltoniana-sigma}
\end{equation}
where the fields ${\bf e}_\alpha$ are $m$ $n$-component vectors 
(or equivalently $m \times n$ matrices) with $n\ge m$ subject to the nonlinear 
constraint
\begin{equation}
{\bf e}_\alpha  \cdot {\bf e}_\beta = \delta_{\alpha\beta} . 
\end{equation}
In order to study the large-$n$ behavior of these models we rewrite the 
general Hamiltonian (\ref{Hgeneral}) in such a way that the dependence 
on the field $\bphi_\alpha$ is quadratic. This is obtained
by introducing two auxiliary fields: a scalar field $S$ and 
a symmetric and
traceless tensor field $T^{\alpha\beta}$,
i.e. such that $T^{\alpha\beta} = T^{\beta\alpha}$, 
$T^{\alpha\alpha} = 0$. By means of these auxiliary fields 
we can rewrite the Hamiltonian (\ref{Hgeneral}) as 
\eq
H = H_{\rm eff} - {3 v_0\over 2} T^2 - 
    {3 w_0\over 2} S^2 + {t_0\over2} A^2 + {1\over 4} F^2,
\label{H-largen}
\en
where 
\eq
H_{\rm eff}  = 
{1 \over 2}\sum_{\alpha,\beta} 
  \bphi_\alpha \cdot X^{\alpha \beta} \bphi_\beta,
\label{Heff}
\end{equation}
and 
\begin{equation}
X^{\alpha \beta} =  - \partial_\mu \partial_\mu \delta^{\alpha \beta} 
  + r_0 \delta^{\alpha \beta}
- 2 g_0 A_\mu^{\alpha \beta} \partial_\mu + 
v_0 T^{\alpha \beta} + w_0 S \delta^{\alpha \beta} + 
    g_0^2 A_\mu^{\beta \gamma} A_\mu^{\alpha \gamma}.
\end{equation}
Note that the effective action for the 
$\phi$ fields is the most general one which is $O(m)$ 
covariant. Therefore, the analysis of this class of models 
provides the critical behavior of the most general 
$O(m)\times O(n)$ theory. 

\section{$\epsilon$-expansion for the Landau-Ginzburg-Wilson 
model} \label{sec-epsilon}

\begin{figure}[tbp]
\centerline{\psfig{figure=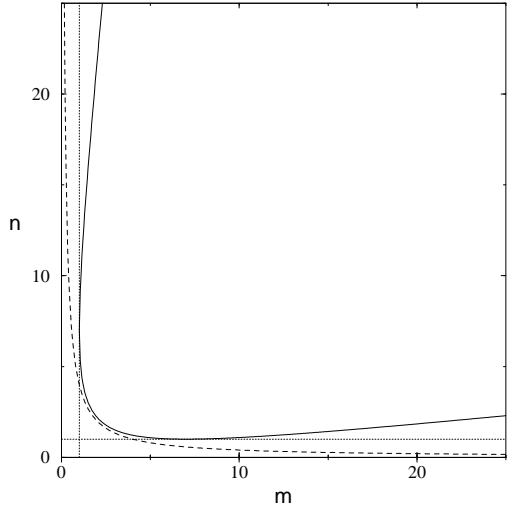,angle=-90,width=0.40\textwidth}}
\caption{The fixed-point structure in the $(m,n)$ plane for $d=4$.
The solid line represents the curves $n^\pm(m)$. The dashed line
shows $n_H(m)$ and the symmetric curve obtained
interchanging $n$ and $m$.
}
\label{fig1}
\end{figure}

In this Section we study the LGW Hamiltonian \reff{H-LGW} and 
report our results for the critical exponents 
and the $\beta$-function to order $\epsilon^3$, thereby extending 
the results of Ref. \cite{Kawa1} to three loops.  
We consider the massless theory and renormalize it using the 
$\overline{\rm MS}$ scheme. We set
\begin{eqnarray}
\bphi &=& [Z_\phi(u,v)]^{1/2} \bphi_R, \\
u_0 &=& \mu^\epsilon Z_u(u,v) N_d^{-1}, \\
v_0 &=& \mu^\epsilon Z_v(u,v) N_d^{-1}, 
\end{eqnarray}
where the renormalization constants are normalized  so that
$Z_\phi(u,v) \approx 1$, $Z_u(u,v) \approx u$, and 
$Z_v(u,v) \approx v$ at tree level. Here $N_d$ is a $d$-dependent constant
given by $N_d^{-1} = 2^{d-1} \pi^{d/2} \Gamma(d/2)$. Moreover,
we introduce a mass renormalization constant $Z_t(u,v)$ by requiring 
$Z_t \Gamma^{(1,2)}$ to be finite when expressed in terms of $u$ and $v$. 
Here $\Gamma^{(1,2)}$ is the two-point function with an insertion of $\phi^2$.
Once the renormalization constants are determined, we compute 
the $\beta$ functions from 
\be
\beta_u (u,v) = \mu \left. {\partial u \over \partial \mu} \right|_{u_0,v_0},
\qquad\qquad
\beta_v (u,v) = \mu \left. {\partial v \over \partial \mu} \right|_{u_0,v_0},
\ee
and the critical exponents from 
\begin{eqnarray}
\eta &=& 
  \left. {\partial \log Z_\phi \over \partial \log \mu} \right|_{u_0,v_0},
\\
\eta_t &=& \left. {\partial \log Z_t \over \partial \log \mu} \right|_{u_0,v_0},
\\
\nu &=& (2 - \eta - \eta_t)^{-1}.
\end{eqnarray}
For the  $\beta$-functions we obtain:\footnote{
We mention that for the particular case $n=m=2$ one may derive
the corresponding four-loop $\epsilon$-expansion from the series reported in
Ref.~\cite{MV-00} for the so-called tetragonal model. Indeed 
an exact mapping \cite{AS-94} exists bringing from the LGW Hamiltonian 
\reff{H-LGW} with $m=n=2$ to the tetragonal model considered in Ref.~\cite{MV-00}.
Note that our renormalized couplings differ from those defined for $m=2$ 
in Ref. \cite{Kawa2}. Kawamura's couplings $u_K$, $v_K$ are related to 
ours by $u = 12 N_d u_K$, $v = 6 N_d v_K$.}
\begin{eqnarray}
\beta_u =&& - \epsilon u +
   {m n +8 \over 6}u^2- {(m-1)(n-1) \over 3}v \left(u-{v \over 2}\right)-
{3 m n +14 \over 12}u^3
\nonumber \\
&&+ (m-1)(n-1)v
   \left({11 \over 18}u^2-{13 \over 24}u v+{5 \over 36}v^2\right)\nonumber \\
&& + {u^4\over1728} 
  \left[ 33 m^2 n^2 + 922 m n + 2960 + 
         \zeta(3) (480 m n + 2112)\right] 
\nonumber \\
&& + {v\over 3456} (m-1)(n-1) \left\{
  -4 [79 m n + 1318 + 768 \zeta(3)] u^3 \right.
\nonumber \\[1mm]
&& \quad \qquad + [555 mn - 460 (m + n) + 6836 + 4032 \zeta(3)] u^2 v 
\nonumber \\[1mm]
&& \quad \qquad - 2 [213 m n - 358 (m+n) + 1933 + 960 \zeta(3)] u v^2
\nonumber \\[1mm]
&& \quad \qquad \left. + [121 mn - 309 (m + n) + 817 + 216 \zeta(3)] v^3
\right\},
\end{eqnarray}
\begin{eqnarray}
\beta_v = &&-\epsilon v + 2 u v + {m+n-8 \over 6}v^2-{5 m n+82 \over 36}u^2 v
\nonumber \\
&&+{5 m n -11(m+n)+53 \over 18}u v^2   -{13 m n -35(m+n)+99 \over 72}v^3
\nonumber \\
&& + {v^4\over 3456} 
   \left\{ 52 m^2n^2 - 57 mn(m+n) - 2206 m n - 111 (m^2 + n^2) + 
   4291 (m+n) - 8084 \right. 
\nonumber \\[1mm]
&& \qquad \qquad 
   \left. \vphantom{(m^2 + n^2)}
   + [- 1416 m n + 3216 (m + n) - 7392] \zeta(3)\right\} 
\nonumber \\
&& + {v^3 u\over 864} \left\{-39 m^2 n^2 + 35 m n (m + n) + 1302 m n +
       36 (m^2 + n^2) - 2401 (m + n) + 5725 \right.
\nonumber \\[1mm]
&& \qquad \qquad
   \left. \vphantom{(m^2 + n^2)}
   + [768 m n - 1824 (m + n) + 4896]\zeta(3)\right\}
\nonumber \\
&& + {u^2 v^2\over 1728} \left\{78 m^2 n^2 - 35 mn(m+n) - 2114 m n + 
     3182 (m+n) - 12520 \right.
\nonumber \\[1mm]
&& \qquad \qquad
   \left. \vphantom{(m^2 + n^2)}
   + [- 1152 m n + 2304 (m+n) - 10368]\zeta(3)\right\}
\nonumber \\
&& + {u^3 v\over 864}\left[
  -13 m^2 n^2 + 368 m n + 3284 + (192 m n + 2688)\zeta(3)\right].
\end{eqnarray}
For the critical exponents we obtain:
\begin{eqnarray}
\eta &=& {mn+2\over 72} u^2 + 
    (m-1)(n-1) v\left( {v\over 48} - {u\over 36}\right) 
   - {(mn+2) (mn+8)\over 1728} u^3 
\nonumber \\
&& + {(m-1)(n-1)\over 3456} v\left[
   v^2 (2 m n - 5 m - 5 n + 26 )\right.
\nonumber \\[1mm]
&& \qquad 
  \left. - 6 u v (mn -m-n+10) + 6 u^2 (mn+8)\right],
\label{eta-epsilon}
\end{eqnarray}
\begin{eqnarray}
{1\over \nu} &=& 2 - {mn+2\over6}u + 
     {(m-1)(n-1)\over 6}v + {5 (mn+2)\over 72} u^2 
\nonumber \\
&& + 5 (m-1)(n-1) v\left( {v\over 48} - {u\over 36}\right) - 
    {(m n+2)(5 m n + 37)\over 288} u^3
\nonumber \\
&& + {(m-1)(n-1)\over 1152} v\left[
   3 v^2 (7 mn - 16 m - 16 n + 79) \right.
\nonumber \\[1mm]
&& \qquad \left. - u v (61 m n - 58 m - 58 n + 550) + 
          12 u^2 (5 m n + 37)\right].
\label{nu-epsilon}
\end{eqnarray}
As discussed at length in Refs. \cite{Kawa2,Kawa1,KawaR}, 
the critical behavior of these systems depends on the values of $n$ and $m$.
In general, the $\beta$-functions admit four solutions: the Gaussian
fixed point ($u^* = v^* = 0$), the $O(mn)$ Heisenberg fixed point
($v^* = 0$) and two new fixed points with nontrivial values of 
$u^*$ and $v^*$, the chiral and the antichiral fixed points. 
These two additional fixed points do not exist for all $n$ and $m$, 
but, at fixed $m$, only for $n \ge n^+(m)$ and $n \le n^-(m)$. 
The functions $n^\pm(m)$ will be computed below. The critical behavior 
depends on the stability of the fixed points. At fixed $m$, for the 
physically relevant case $m\ge 1$, 
the $\epsilon$-expansion predicts four regimes:
\begin{itemize}
\item[1)] For $n>n^+(m)$, there are four fixed points, and the chiral one 
is the stable one.
\item[2)] For $n^-(m) < n < n^+(m)$, only the Gaussian and the Heisenberg
O($n\times m$)-symmetric fixed points are present, and none of them is stable.
\item[3)] For $n_H(m) < n < n^-(m)$, there are again four fixed points,
and the chiral one is the stable one. 
For small $\epsilon$, the chiral fixed point has $v < 0$ for $m<7$ and 
$v > 0$ for $m > 7$.
\item[4)] For $n < n_H(m)$, there are again four fixed points,
and the Heisenberg O$(m\times n)$-symmetric one is the stable one. 
\end{itemize}
The antichiral fixed point is Gaussian 
for $m\to 1$ and $m=-2$ (or, equivalently for $n\to 1$ and $n=-2$). 
Indeed, for $m\to 1$, $u^*\to 0$ for the antichiral
fixed point, so that, from Eqs. \reff{eta-epsilon} and \reff{nu-epsilon}, 
we obtain $\eta=0$ and $\nu = 1/2$ at order $\epsilon^3$. For $m=-2$
we obtain $u^* = 3v^*/2$ and again $\eta=0$ and $\nu = 1/2$ at order
$\epsilon^3$. We conjecture that this holds to all orders in $\epsilon$, 
and in the next Section we will provide a large-$n$ interpretation of these 
results. The general behavior for $n$ and $m$ is better understood 
from Fig. \ref{fig1}. In particular, the two functions $n^\pm(m)$ are 
nothing but the two different branches of the curve that separates the 
region in which no fixed point is stable from the region in which 
the chiral fixed point is the stable one. Note that the boundaries of the 
different regions are symmetric under the exchange $(n,m)$. Because of this 
symmetry it is more natural to consider the behavior in the variables
\be
\Sigma = m + n, \qquad\qquad \Delta = m - n.
\ee
At fixed $\Delta$ there are then three regions:
\begin{itemize}
\item[1)] For $\Sigma > \Sigma^+$ only the Gaussian and the Heisenberg
O($m\times n$)-symmetric fixed points
are present and none is stable;
\item[2)] For $\Sigma_H < \Sigma \le \Sigma^+$ there are four fixed point and the 
chiral one is stable;
\item[3)] For $\Sigma \le \Sigma_H$, there are four fixed points and
the  O$(m\times n)$-symmetric one is the stable one.
\end{itemize}

Using the above results, we can compute the $\epsilon$ expansion 
of $n^\pm(m)$ and $n_H(m)$. For $n^\pm(m)$ we expand
\be
n^\pm(m) = n_0^\pm + n_1^\pm \epsilon + n_2^\pm \epsilon^2 + 
          O(\epsilon^3).
\ee
Then, by requiring 
\be
\beta_u(u^*,v^*;n^\pm) = 0, \qquad \qquad
\beta_v(u^*,v^*;n^\pm) = 0, 
\ee
and 
\be 
{\rm det}\, \left|{\partial(\beta_u,\beta_v)\over \partial(u,v)} \right| 
 (u^*,v^*;n^\pm) = 0,
\ee
we obtain
\begin{eqnarray}
n_0^\pm &=& 5 m + 2 \pm 2 s, \\
n_1^\pm &=& - 5 m - 2 \mp {1\over 2s} (25 m^2 + 22 m - 32), \\
n_2^\pm &=& {1\over24} {R_1\over Q_1} \pm {1\over 32} {R_2\over s Q_1} + 
            {1\over8} {R_3\over Q_2} \zeta(3) \pm 
            {1\over 48} {s R_4\over Q_2} \zeta(3).
\end{eqnarray}
Here
\begin{eqnarray}
s &=& \sqrt{6 (m-1) (m+2)}, \nonumber \\
R_1 &=& -33024 + 18880 m + 45444 m^2 + 9288 m^3 - 1883 m^4 - 417 m^5 +
      21 m^6 + 4 m^7, \nonumber \\
R_2 &=& -253952 - 160256 m + 176192 m^2 + 139240 m^3 + 7756 m^4 - 5854 m^5 
\nonumber \\
&& \qquad
        - 389 m^6 + 58 m^7 + 5 m^8, \nonumber \\
R_3 &=& 1632 + 1184 m - 1376 m^2 - 426 m^3 + 31 m^4 + 8 m^5, \nonumber \\
R_4 &=& 6176 - 2960 m - 1230 m^2 + 73 m^3 + 20 m^4, \nonumber \\
Q_1 &=& (m+8)^2 (m+2) (m-1) (m-7)^2, \nonumber \\
Q_2 &=& (m+8) (m+2) (m-1) (m-7). 
\end{eqnarray}
For $m=2$ this expression is in agreement with that given in Ref. \cite{ASV-95}.
The expression for $n_2^\pm$ are singular for $m=7$. 
However, this is not the case for $n_2^+$, and indeed, by taking the limit 
we obtain
\be
n_2^+ = {23871617\over 9331200} + {5487\over 320} \zeta(3).
\ee
For $n_H(m)$ we have
\be 
n_H(m) = {1\over m}\left[4 - 2 \epsilon + \case{5}{12}(6 \zeta(3) - 1)
       \epsilon^2 + O(\epsilon^3)\right],
\ee
which is a trivial generalization of the result of Ref. \cite{ASV-95}.
The calculation of the functions $\Sigma^+(\Delta)$ and $\Sigma_H(\Delta)$ 
follows the same lines. 
In particular,
\be
\Sigma^+(\Delta) = - 1 + {1\over2}\tilde{s} + {\epsilon\over 8\tilde s}
        \left(5 \Delta^2 - 24 - 2 \tilde{s}\right) + O(\epsilon^2),
\ee
where $\tilde{s} = \sqrt{6 (\Delta^2 + 18)}$.

From our calculation of the RG functions we can derive the 
fixed points of the theory. We expand 
\begin{eqnarray}
&&u^{\ast} = u_1 \epsilon+ u_2 \epsilon^2 + 
    u_3 \epsilon^3 + O(\epsilon^4) ,
\nonumber \\
&&v^{\ast} = v_1 \epsilon + v_2 \epsilon^2 + 
    v_3 \epsilon^3 + O(\epsilon^4).
\end{eqnarray}
Following Ref. \cite{Kawa1} we define
\begin{eqnarray}
&&B_{mn}^{-1} \equiv (m n+8)(m+n-8)^2+ 24(m-1)(n-1)(m+n-2), \\
&& D_{mn} \equiv m n(m+n)-10(m+n)+4 m n -4, \nonumber \\
&& R_{mn} \equiv (m+n-8)^2-12(m-1)(n-1). \nonumber
\end{eqnarray}
We then easily find:
\begin{eqnarray}
&&u_1^{\pm} = {1 \over 2}- {1 \over 2}(m+n-8)B_{mn}
   \left[D_{mn}\mp 6 R_{mn}^{1 /2} \right], \nonumber \\
&&v_1^{\pm} = 6 B_{mn} \left[ D_{mn} \mp 6 R_{mn}^{1 /2} \right],
\end{eqnarray}
where we indicate by $(+)$ the stable chiral fixed point and by 
$(-)$ the unstable antichiral one. In order to compute 
$u_2^\pm$ and $v_2^\pm$, it is convenient to define two additional 
auxiliary functions:
\begin{eqnarray}
&&S_1 \equiv -{(3 m n+14) \over 12}u_1^3 + (m-1)(n-1) v_1
  \left({11 \over 18} u_1^2- {13 \over 24}u_1 v_1+ {5 \over 36} v_1^2\right),
\\
&&S_2 \equiv -{(5 m n+82) \over 36}u_1^2 v_1+
    {5 m n-11(m+n)+53 \over 18}u_1 v_1^2-{13 m n -35(m+n)+99 \over 72}v_1^3.
\nonumber
\end{eqnarray}
Then,
the $O(\epsilon^2)$ coefficients of the fixed points are given by
\begin{eqnarray}
&&u_2^{\pm} = \pm 6 
   {\left(1-2 u_1^{\pm}-{m+n-8 \over 3}v_1^{\pm}\right)S_1^\pm 
   - {(m-1)(n-1) \over 3}
   \left(u_1^{\pm}- v_1^{\pm}\right)S_2^\pm 
      \over v_1^{\pm} R_{mn}^{1 /2}},
\nonumber \\
&&v_2^{\pm} = \pm 6 {2 v_1^{\pm} S_1^\pm+
   \left(1- {m n+8 \over 3}u_1^{\pm}+{(m-1)(n-1) \over 3}v_1^{\pm}\right)
S_2^\pm \over v_1^{\pm} R_{mn}^{1/  2}}. 
\end{eqnarray}
The expressions for $u_3^{\pm}$, $v_3^{\pm}$ are particularly cumbersome and 
they will not be reported here. 

Once the fixed points are determined, the critical exponents 
are computed by expanding in power of $\epsilon$ the exponent series 
computed at the fixed point. 
Such a computation gives us the exponents only for 
\be
    n > 5 m + 2 + 2 \sqrt{6 (m+2)(m-1)},
\ee
or 
\be
    n < 5 m + 2 - 2 \sqrt{6 (m+2)(m-1)}.
\ee
Indeed, if these bounds are not satisfied the fixed points are complex 
and therefore also the exponent series. In order to 
obtain series for the exponents in all the relevant domain we can perform the 
following trick. For $n > n^+$, which is the case of physical interest,
we set $n = n^+(m,\epsilon) + \Delta n$ and reexpand all series in powers of 
$\epsilon$ keeping $\Delta n$ fixed. In particular, for 
$\Delta n = 0$ we obtain the critical exponents for $n = n^+$. 
In such a case, for $m=2$ we obtain:
\begin{eqnarray}
&&\eta = {1\over48}\epsilon^2 + {5\over288}\epsilon^3,
\nonumber \\
&& {1\over\nu} = 2 - {1\over2}\epsilon + 
    \epsilon^2\left({\sqrt{6}\over50} - {1\over50}\right)
    + \epsilon^3\left({397\over15000} - {37\sqrt{6}\over 15000} + 
     {37\over 1000}\zeta(3) + {\sqrt{6}\over250}\zeta(3)\right).
\label{exponents-ncritico}
\end{eqnarray}

\section{The Landau-Ginzburg-Wilson theory in the 
large-$n$ limit} \label{sec-largen}

In this Section, we study the large $n$-behavior of the LGW theory
(\ref{H-LGW}) at fixed $m$. The starting point is the general Hamiltonian 
(\ref{H-largen}) with $A_\mu^{\alpha\beta} = 0$. 
In the high-temperature phase the symmetry is unbroken and thus 
the relevant saddle point is given by 
\begin{eqnarray}
\langle S\rangle = \sigma, \qquad\qquad 
\langle T_{\alpha\beta} \rangle = 0 .
\end{eqnarray}
Correspondingly, we obtain the gap equation:
\begin{eqnarray}
\int {d^dp\over (2\pi)^d} {1\over p^2 + M^2} = {6 \sigma\over n m},
\label{gap-equation1}
\end{eqnarray}
where 
$M^2 = r_0 + {w_0\sigma}$. For the $\sigma$-model \reff{sigma-model-gauged}
we obtain analogously\footnote{This result can be obtained by using  
Eq. \reff{gap-equation1} and by taking the limit considered 
before Eq. \reff{sigma-model-gauged}. Notice 
that, in order to keep $M^2$ finite in the limit, $\sigma$ must 
converge to $m \eta_1$ as $v_0\to\infty$.}
\be
\int {d^dp\over (2\pi)^d} {1\over p^2 + M^2} = {\eta_1\over n}.
\ee
From the gap equation we can obtain the scaling of the mass and thus 
the exponent $\nu$. For $w_0\not = 0$, proceeding as in the case of the 
ordinary $O(n)$ model, we obtain for $2<d<4$
\be
   \nu = {1\over d-2}.
\label{nu-ninf-1}
\ee
However, for $w_0 = 0$, we obtain simply $M^2 = r_0$, indicating
\be
   \nu = {1\over 2}
\label{nu-ninf-2}
\ee
for all values of the dimension $d$.

Within the large-$n$ limit 
we can recover the critical behavior of the theory at all fixed 
points. For generic $v_0$ and $u_0$, satisfying $w > 0$, $v_0 > 0$ 
we obtain the critical chiral behavior. The standard Heisenberg behavior 
is obtained by setting $v_0 = 0$, while the 
antichiral critical behavior is obtained at the stability boundary,
i.e. by setting $w_0 = 0$. It is easy to see the different 
types of excitations that appear in these cases: 
at the chiral fixed point both the scalar and the tensor degrees 
of freedom propagate, while at the Heisenberg and antichiral 
fixed point one observes only the scalar and the tensor degrees of 
freedom respectively. Note that, as a consequence of 
Eqs. \reff{nu-ninf-1} and \reff{nu-ninf-2}, the Heisenberg and the 
chiral point have the same exponents for $n=\infty$, and that 
they differ from those of the antichiral point which shows 
mean-field behavior in all dimensions.

In order to perform the calculation we heavily relied upon the results 
obtained by Vasil'ev {\it et al.}
\cite{Vasi1,Vasi2}, who studied the models corresponding to 
$m=1$ with a method which lends itself to a reasonably simple extension, 
appropriate to the case we are investigating.
In order to make our presentation self-contained, 
we must briefly review the essentials of the method.

\begin{figure}[t]
\vskip 1truecm
\centerline{\psfig{width=10truecm,angle=0,file=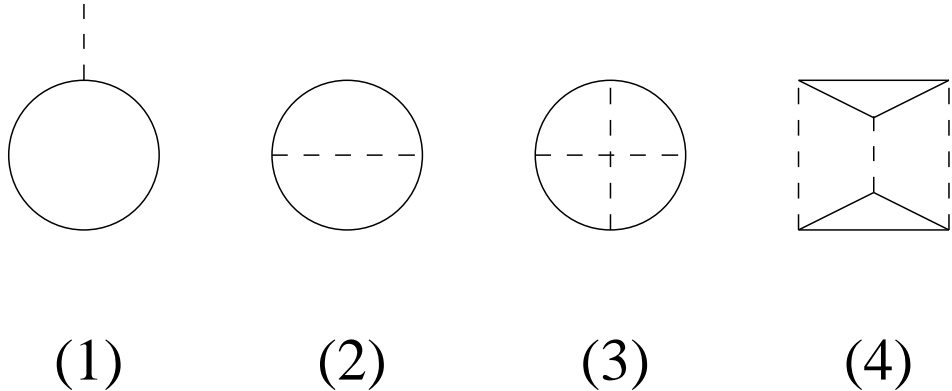}}
\vskip -1truecm
\caption{The graphs appearing in second Legendre transform.
The continuous line represents the dressed propagator of the
$\phi$ field, while the dashed line indicates the dressed propagator
of an auxiliary field.}
\label{fig-vacuum}
\end{figure}

One first considers the second Legendre transform with respect to the 
field and the two-point function \cite{Cornwall-etal}. We indicate by 
$D^{\alpha a,\beta b}_\phi (p)$, $D_S(p)$, and 
$D^{\alpha\beta,\gamma\delta}_T(p)$ the dressed propagators of the 
field $\phi^{\alpha a}$ and of the auxiliary fields $S$ and $T^{\alpha\beta}$.
Here $\alpha$, $\beta$, $\gamma$, and $\delta$ go from 1 to $m$, 
while $a$ and $b$ go from 1 to $n$.
It is useful to factorize the group dependence and to introduce 
scalar propagators 
\begin{eqnarray}
D^{\alpha a,\beta b}_\phi (p) &=& \delta^{\alpha\beta} \delta^{ab} 
    \hat{D}_\phi (p), 
\label{Dphi} \\
D_T^{\alpha\beta,\gamma\delta}(p) &=& 
  \case{1}{2} \left(
    \delta^{\alpha\gamma} \delta^{\beta\delta} + 
    \delta^{\alpha\delta} \delta^{\beta\gamma} -
  {2\over m} \delta^{\alpha\beta} \delta^{\gamma\delta} \right) 
   \hat{D}_T (p).
\label{DT}
\end{eqnarray}
Also we can reabsorb the coupling $v_0$ and $w_0$ in the fields. 
Using the same notations of Ref. \cite{Vasi1}, we have for the 
second Legendre transform
\begin{eqnarray} 
\Gamma &=& \case{1}{2} {\rm Tr}\, \log D_\phi + 
         \case{1}{2} {\rm Tr}\, \log D_S    +
         \case{1}{2} {\rm Tr}\, \log D_T    +
         \case{n}{2} (\gamma_1(S) + \gamma_1(T)) + 
\vphantom{1\over2} \nonumber \\
&& \case{1}{4}{n m} \gamma_2(S) + \case{1}{8}n (m-1) (m+2) \gamma_2(T) + 
   \case{1}{8} {n m} \gamma_3(SS) + \case{1}{8} n (m-1) (m+2) \gamma_3(TS) + 
\vphantom{1\over2} \nonumber \\
&& {n\over 32 m} (m-1) (m^2 - 4) \gamma_3(TT) + 
   \case{1}{12} n^2 m^2 \gamma_4(SSS) + 
\nonumber \\
&& \case{1}{8} n^2 (m-1) (m+2) (\gamma_4(SST) + \gamma_4(STT)) + 
   {n^2\over 96 m} (m-1) (m^2 - 4) (m+4) \gamma_4(TTT) + \ldots
\nonumber \\
{} \label{second-Legendre-transform}
\end{eqnarray}
In this equation, $\gamma_1$, $\ldots$, $\gamma_4$ are the graphs reported 
in Fig. \ref{fig-vacuum} and the letters in parentheses indicate which 
auxiliary fields are propagating in the graph. In these graphs one 
should use the scalar dressed propagators $\hat{D}_\phi$ and $\hat{D}_T$ 
while each vertex is trivially one. 
In the equation we have of course reported only those graphs that are 
relevant for the computation of the critical indices at order $1/n^2$.
The group-theoretical factors in Eq. \reff{second-Legendre-transform} 
have been obtained by using Eqs. \reff{Dphi} and \reff{DT} and keeping into 
account that the vertex $\phi\phi T$ has the form: 
\be
\phi_{\alpha a} \phi_{\beta b} T^{\gamma\delta} \to 
\case{1}{2} \delta^{ab} \left( \delta^{\alpha \gamma} \delta^{\beta\delta}
   + \delta^{\alpha \delta} \delta^{\beta\gamma} - 
   {2\over m} \delta^{\alpha\beta}\delta^{\gamma\delta}\right).
\ee
Eq. \reff{second-Legendre-transform} is completely general and can be used in
the computation of the critical indices for all fixed points: while 
we should keep into account all terms for the chiral fixed point, 
we should set $T=0$ and $S=0$ for the Heisenberg and the antichiral fixed 
points respectively.

\begin{figure}[t]
\vskip 1truecm
\centerline{\psfig{width=10truecm,angle=0,file=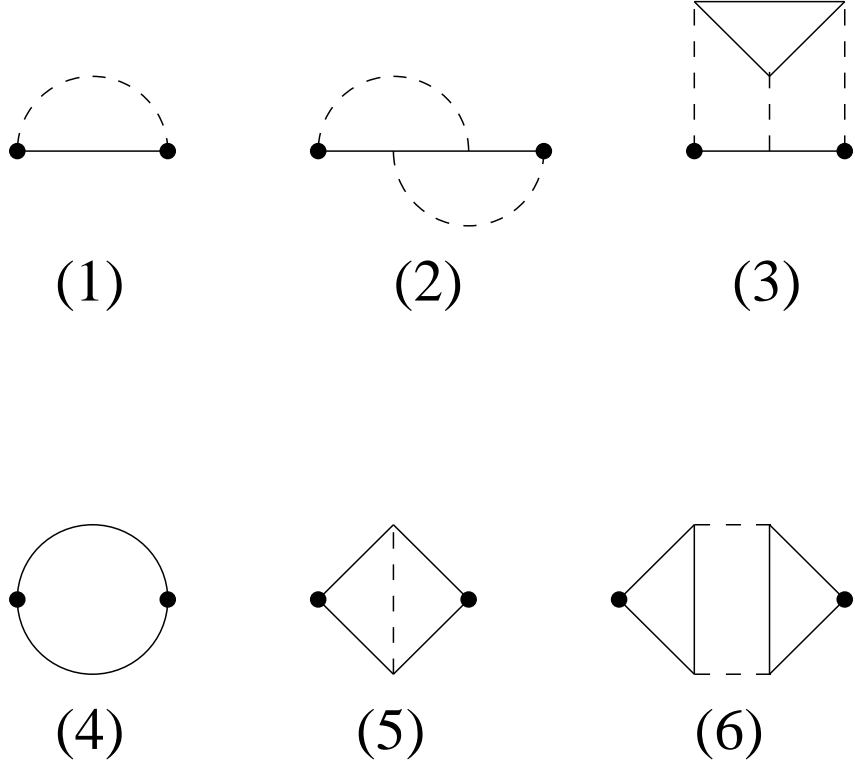}}
\vskip -1truecm
\caption{The graphs appearing in the Dyson equations. 
The continuous line represents the dressed propagator of the 
$\phi$ field, while the dashed line indicates the dressed propagator 
of an auxiliary field.}
\label{fig-2pt}
\end{figure}

From Eq. \reff{second-Legendre-transform} we can derive the skeleton 
Dyson equations for the dressed propagators. It is enough to 
compute the variation of $\Gamma$ with respect to the dressed 
propagators. We obtain for the field $\phi$ the equation
\begin{eqnarray}
&& \hat{D}_\phi^{-1} - \Delta + u + g_1(S) + 
   {1\over 2m} (m-1)(m+2) g_1(T) + 
\nonumber \\
&& \quad
   g_2(SS) + {1\over m} (m-1)(m+2) g_{2}(ST) + 
   {1\over 4 m^2} (m-1) (m^2 - 4) g_2(TT) 
\nonumber \\
&& \quad + mn g_{3}(SS) + 
    {n\over 2m}(m-1)(m+2) (g_3(STS) + 2 g_3(TSS) + g_3(TST) + 2 g_3(TTS))
\nonumber \\
&& \quad {n\over 8 m^2} (m-1) (m^2 - 4) (m+4) g_3(TTT) + \ldots = 0,
\label{skeleton-1}
\end{eqnarray}
while for the auxiliary fields we have
\begin{eqnarray}
&& D_S^{-1} + c_S {nm\over2} g_4 + nm g_5(S) + 
  {n\over2} (m+2)(m-1) g_5(T) + 
\nonumber \\
&& \quad + {n^2 m^2 \over2} g_6(SS) + 
   {n^2\over2} (m+2)(m-1) g_6(TT) + \ldots = 0;
\label{skeleton-2}
\end{eqnarray}
\begin{eqnarray}
&& \hat{D}_T^{-1} + c_T + {n\over2} g_4 + {n\over2} g_5(S) + 
    {n\over2m} (m-2) g_5(T) + 
\nonumber \\
&& \quad + n^2 g_6(ST) + {n^2\over8m} (m+2)(m-4) g_6(TT) + \ldots = 0.
\label{skeleton-3}
\end{eqnarray}
Here $c_S$ and $c_T$ are two constants, $u$ a momentum-independent 
contribution due to the tadpoles, $\Delta = p^2 + M^2$, 
and $g_i$ are the graphs reported in Fig. \ref{fig-2pt}. As before, 
in parentheses we report which auxiliary fields are propagating. 
Each line is associated to a scalar dressed propagator, while the vertices 
are one. 

The critical exponents are determined following closely the method of Ref. 
\cite{Vasi1}. As in Ref. \cite{Vasi1} we introduce two auxiliary functions:
\begin{eqnarray}
&&a(x) \equiv {\Gamma({d \over 2}-x) \over \Gamma(x)},\;\;\;\;p(x) \equiv {a(x-{
d \over 2})
\over \pi^d a(x)}, \\
&&q(x,y) \equiv {a(x-y) a(x+y-{d \over 2}) \over a(x) a(x-{d \over 2})}.
\end{eqnarray}
Note the trivial symmetry of the function $q(x,y)$ which will play a 
role below: 
\be
q(x,y) = q(x,\case{d}{2} - y).
\label{q-symmetry}
\ee
The calculation of the $1/n$ correction 
starts from assuming for $x\to 0$ the following 
behavior of the dressed propagators:
\be
    D_X(x) = {A_X\over x^{2 \alpha_X}} (1 + B_X x^{2\lambda} + \ldots),
\label{defDX}
\ee
where $X$ is any of the fields. Here $\alpha_X$ is related to the 
dimension of the field. For $X=\phi$ we have
$\alpha_\phi = d/2 - 1 + \eta/2$. The correction
term we report is the analytic one in the temperature and therefore 
$\nu = 1/(2 \lambda)$. From Eq. \reff{defDX} we obtain for the inverse 
functions: 
\be
D_X^{-1}(x) = {p(\alpha_X)\over A_X x^{2 d - 2 \alpha_X} }
    \left[1 - B_X q(\alpha_X,\lambda) x^{2\lambda} + \ldots\right].
\ee
Plugging these expressions in the skeleton equations and equating the 
corresponding terms we obtain six equations for the amplitudes. 
Such equations have nontrivial solutions only if 
\be
\alpha_S = \alpha_T = 2 - \eta \equiv \beta,
\ee
and the following consistency equations are satisfied: 
\begin{eqnarray}
&&p(\alpha_\phi)=2 {M \over n}p(\beta),
\label{consistency-1} \\
&&[q(\alpha_\phi,\lambda)+1]q(\beta,\lambda)=2,
\label{consistency-2}
\end{eqnarray}
where $M$ is a group-theoretical factor that depends on the fixed point:
\be
M = \, \cases{
    M^+ \equiv  \case{1}{2}(m+1) 
     \vphantom{\displaystyle{1\over2}} & \qquad\qquad {\rm (chiral f.p.)}; \cr
    M^- \equiv \displaystyle{1\over 2m}(m-1)(m+2) 
     \vphantom{\displaystyle{1\over (2^2)^2}} & \qquad\qquad {\rm (antichiral f.p.)};
    \cr
    M^H \equiv  \displaystyle{1\over m} \vphantom{\displaystyle{1\over(2^2)^2}} 
    & \qquad\qquad {\rm (Heisenberg f.p.)}. 
    }
\label{defM}
\ee
Note that $M^-=0$ for $m=1,-2$, a result which follows from the fact 
that only a symmetric traceless tensor propagates.
Eq. \reff{consistency-1} allows the determination of the first 
large-$n$ coefficient appearing in the expansion of the exponent $\eta$,
\be
\eta = {\eta_1\over n} + {\eta_2\over n^2} + {\eta_3\over n^3} + \ldots 
\ee
For $\eta_1$ we obtain 
\begin{equation}
\eta_1= M \eta_{11},
\end{equation}
where
\begin{equation}
\eta_{11}= - {4\Gamma(d-2) \over \Gamma(2-{d \over 2})\Gamma({d \over 2}-1)
\Gamma({d \over 2}-2)\Gamma({d \over 2}+1)},
\end{equation}
and the dependence on the fixed point is encoded in the factor $M$.
The quantity $\eta_{11}$ is the well-known result for the $m=1$ model; among
its most important properties we wish to mention that 
$\eta_{11} \rightarrow 0$ both in the $d \rightarrow 4$ and 
in the $d \rightarrow 2$ limit.

Equation \reff{consistency-2} should allow the determination of the 
exponent $\nu$. However, there is a subtle point that has been overlooked
in the previous analyses. 
It is convenient for our discussion to introduce the auxiliary function
\begin{equation}
r(\eta,\lambda) \equiv \left[q\left({d \over 2}+{\eta \over 2}-1, 
            \lambda\right)+1\right]
q(2- \eta, \lambda),
\end{equation}
which corresponds to the left-hand side of Eq.
\reff{consistency-2}. Because of Eq. \reff{q-symmetry}, the function 
$r(\eta,\lambda)$ has the symmetry 
\begin{equation}
r(\eta,\lambda) \equiv r(\eta, \case{d}{2}-\lambda).
\end{equation}
As a consequence, once $\eta$ is fixed by solving 
Eq. \reff{consistency-1}, 
we still have the possibility of finding {\it two} different solutions
for $\lambda\equiv 1/2\nu$. 
It is convenient to parametrize the two solutions by
\begin{equation}
\lambda \equiv {d \over 4}\pm\left({d \over 4}-1+\rho\right).
\label{defrho}
\end{equation}
Using Eq. \reff{consistency-2} and the fact that $\eta$ is of order 
$1/n$, one can show that also $\rho$ is of order $1/n$ and therefore 
has an expansion of the form
\be
\rho = {\rho_1\over n} + {\rho_2\over n^2} + \ldots
\ee
The coefficient $\rho_1$ is computed from Eq. \reff{consistency-2} 
using the fact that for $n$ large
\be
q\left(\case{d}{2} - 1 + \case{1}{2}\eta, \case{d}{2} - 1 + \rho\right) = 
q\left(\case{d}{2} - 1 + \case{1}{2}\eta, 1 - \rho\right) = 
   {4-d \over d}\left(1-{2\rho_1 \over \eta_1}\right) + O(n^{-1}).
\end{equation}
We then obtain
\be 
\rho_1 = M\rho_{11},
\end{equation}
where
\begin{equation}
\rho_{11} = {(d-1)(d-2) \over 4-d}\eta_{11}.
\end{equation}
As we observed the consistency equations are satisfied for two independent
choices of $\lambda$. In order to associate to correct 
one to each fixed point we use the large-$n$ estimates of the exponent 
$\nu$ given in Eqs. \reff{nu-ninf-1} and \reff{nu-ninf-2}. 
Then we have 
\be
\lambda = \cases{\displaystyle{{d\over2} - 1 + {1\over n} M^+ \rho_{11}  }
   & \quad chiral f.p.; \cr
     \displaystyle{    1 - {1\over n} M^- \rho_{11} }
   & \quad antichiral f.p.; \cr
      \displaystyle{  {d\over2} - 1 + {1\over n} M^H \rho_{11} }
   & \quad Heisenberg f.p.
}
\ee
The corrections of order $1/n^2$ can be obtained by generalizing 
the arguments of Refs. \cite{Vasi1,Vasi2}.
One must only pay attention to insert proper group-theoretical
factors in front of the corresponding $m=1$ contributions: they can be obtained
from Eqs. \reff{skeleton-1}, \reff{skeleton-2}, and 
\reff{skeleton-3}. 

We introduce the following definitions: 
\begin{eqnarray}
&&\eta_{21}^{HF}= (\eta_{11})^2
   \left[\Psi+{d^2+2 d-4 \over 2 d(d-2)}\right],\\
&&\rho_{21}^{HF}= (\eta_{11})^2{d-1 \over (d-4)^2}
   \left[(d-2)(4+2 d-d^2)\Psi+{32 +
         8 d-30 d^2+7 d^3 \over d(d-4)}\right],
\end{eqnarray}
where
\begin{equation}
\Psi \equiv \psi(d-2)+
            \psi\left(2- {d \over 2}\right)-
            \psi\left({d \over 2}-2\right)-\psi(2).
\end{equation}
Also:
\begin{eqnarray}
\eta_{21}^{(a)}= &&(\eta_{11})^2
   \left({d \over 4-d}\Psi+ {d(6-d) \over 2 (4-d)^2}\right),\\
\eta_{21}^{(b)}= &&(\eta_{11})^2
   \left({d(d-3) \over 4-d}\Psi+{d(d-2) \over 4-d}\right),
\nonumber \\
\rho_{21}^{(a)}=  &&{d(d-2) \over 2(4-d)^2} (\eta_{11})^2 
   \left[2(d-1) \Psi+{3 \over 2}d(d-3)R_1+2d-8
+{6 \over 4-d}-{4 \over d-2}+{12 \over (d-2)^2} \right] ,\nonumber \\
\rho_{21}^{(b)}=  &&{d(d-2) \over 2(4-d)^2} (\eta_{11})^2 
\left[{4(d-3) \over (d-2)} \Psi+{3 \over 2}d(3d-8)R_1 \right.
\nonumber\\
&& \left. +{2d(d-3)^2 \over (4-d)}(6 R_1-R_2-R_3^2)+d^2+d+20-
   {12(4-d) \over(d-2)^2}-{16 \over 4-d}\right] ,
\end{eqnarray}
where
\begin{eqnarray}
&&R_1\equiv \psi^\prime\left({d \over 2}-1\right)-\psi^\prime(1),
\nonumber \\
&&R_2 \equiv \psi^\prime(d-3)-
    \psi^\prime\left(2-{d \over 2}\right)-
    \psi^\prime\left({d \over 2}-1\right)+
    \psi^\prime(1),\nonumber\\
&&R_3 \equiv \psi(d-3)+\psi\left(2-{d \over 2}\right)-
             \psi\left({d \over 2}-1\right)-\psi(1).
\end{eqnarray}
Recalling the above considerations about the choice of $\lambda$ in conjunction
with the choice of the critical point, 
we can now write down our final results for the $1/n$
expansion of the critical exponents in the  $O(n) \times O(m)$ models, 
both for the chiral (stable) critical point and 
for the antichiral (unstable) one:
\begin{eqnarray}
\eta^+ = &&
  {m+1 \over 2 n}\eta_{11}+ {1 \over n^2}
   \left[{(m+1)^2 \over 4} \eta_{21}^{HF}+
      {m+3 \over 4}\eta_{21}^{(a)}+ {m^2+3m+4 \over 8}\eta_{21}^{(b)}\right] +
   O\left({1 \over n^3}\right), 
   \hphantom{pppp}
\label{etapiu-largen}\\
\eta^- = &&
  {(m-1)(m+2) \over 2 m n}\eta_{11}+ {1 \over n^2}
     \left[{(m-1)^2 (m+2)^2 \over 4 m^2}
      \eta_{21}^{HF}+ {(m-1)(m^2-4) \over 4 m^2}\eta_{21}^{(a)}
\right. \nonumber \\
&& \left. + {(m-1)(m^2-4)(m+4) \over 8 m^2}\eta_{21}^{(b)}\right] +
    O\left({1 \over n^3}\right),\\
\nu^+ = &&{1 \over d-2}-{2 \over (d-2)^2}{m+1 \over 2 n}\rho_{11}- 
    {2 \over (d-2)^2}{1 \over n^2}
    \left[{(m+1)^2 \over 4}\rho_{21}^{HF}+ {m+3 \over 4}\rho_{21}^{(a)}
\right. \nonumber \\
&& \left. +{m^2+3m+4 \over 8}\rho_{21}^{(b)}
-{2 \over d-2}{(m+1)^2 \over 4}\rho_{11}^2\right]+
   O\left({1 \over n^3}\right),
\label{nupiu-largen} \\
\nu^- = &&{1 \over 2}+{1 \over 2}{(m-1)(m+2) \over 2 m n}\rho_{11}+ 
   {1 \over 2}{1 \over n^2}
   \left[{(m-1)^2 (m+2)^2 \over 4 m^2}\rho_{21}^{HF}
\right. \nonumber \\
&& + {(m-1)(m^2-4) \over 4 m^2}\rho_{21}^{(a)}
+{(m-1)(m^2-4)(m+4) \over 8 m^2}\rho_{21}^{(b)}\nonumber\\
&&\left. +{(m-1)^2 (m+2)^2 \over 4 m^2}\rho_{11}^2 \right]
+O\left({1 \over n^3}\right).
\label{numeno-largen}
\end{eqnarray}
The expressions for the stable fixed point at order $1/n$ coincide with those 
of Ref. \cite{Kawa3}. Note that $\eta^-=0$, $\nu^- = 1/2$ for 
$m=-2,1$, in agreement with our $\epsilon$-expansion results.

It is possible to expand the above large-$n$ results in powers of 
$\epsilon= 4-d$.
The resulting expressions can be compared with the 
the $\epsilon$-expansion
results for the LGW hamiltonian presented in the previous Section.
We find full agreement both for the stable and 
the unstable fixed point 
for all $m$, thus confirming our identification of the large-$n$ fixed 
points with the perturbative ones.

For $d=3$ the large-$n$ expansions simplify to:
\begin{eqnarray}
\eta^+ &=& {4(m+1)\over 3 n \pi^2}  + 
  {16 (m^2 - 7 m - 26)\over 27 n^2 \pi^4} + 
       O\left({1 \over n^3}\right), \\
\eta^- &=& {4(m-1)(m+2)\over 3 m n \pi^2} + 
   {16 (m-1)(m+2)(m^2-8m-2)\over 27 m^2 n^2 \pi^4} + 
     O\left({1 \over n^3}\right), \\
\nu^+ &=& 1 - {16 (m+1)\over 3 n \pi^2} 
         - {1\over n^2} \left({4 (m^2 + 3 m + 4)\over \pi^2} - 
         {64 (5 m^2 + 19 m + 32)\over 27 \pi^4}\right) + 
      O\left({1 \over n^3}\right), \\
\nu^- &=& {1\over2} + {4 (m+2)(m-1)\over 3 m n \pi^2} \nonumber \\
&& + 
 {(m-1)(m+2)\over 27 m^2 n^2\pi^4} 
 \left[ 16(13m^2 + 4 m + 28) + 27 (m+4) (m-2) \pi^2\right] + 
  O\left({1 \over n^3}\right). \hphantom{ppppp}
\end{eqnarray}

\section{ The $1/n$-expansion in the presence of a vector field} 
\label{sec-vector}

It is quite instructive to extend the discussion of the 
previous paragraph to the 
more general case in which gauge-invariant vector degrees of freedom are 
allowed. 
This corresponds to studying the general Lagrangian \reff{Hgeneral}
with $t_0=0$.

In the large-$n$ limit one starts from Eqs. \reff{H-largen} and 
\reff{Heff} with $t_0=0$. As discussed by Hikami \cite{Hika}, the gauge 
kinetic term is irrelevant for $d<4$, as well as the $T^2$ and $S^2$ terms 
in Eq. \reff{H-largen}. Thus, the large-$n$ limit can be studied 
by keeping only into account $H_{\rm eff}$. The discussion of the fixed points 
is identical to that presented in the previous Section. If $g_0\not=0$ 
a new set of fixed points appear: for generic $v_0>0$, $w_0>0$ 
we have the chiral-gauge fixed point in which all excitations 
($S$, $T^{\alpha\beta}$, $A^{\alpha\beta}_\mu$) propagate; for 
$w_0 = 0$ we have the antichiral-gauge fixed point ($T^{\alpha\beta}$,
$A^{\alpha\beta}_\mu$), for $v_0= 0$ the Heisenberg-gauge fixed point 
($S$, $A^{\alpha\beta}_\mu$), and for $u_0=v_0=w_0=0$ the pure gauge 
fixed point ($A^{\alpha\beta}_\mu$). 

Here, we want to compute the critical behavior for $n\to\infty$,
keeping only the leading correction. Since the model is gauge-invariant,
the large-$n$
propagator of the field $A^{\alpha\beta}_\mu$ is not uniquely defined. 
Indeed, by integration over the field $\bphi$ we obtain a 
coupling ${1\over2} A^{\alpha\beta}_\mu A^{\gamma\delta}_\nu 
M^{\alpha\beta,\gamma\delta}_{\mu\nu}$, where in momentum space
\be
M^{\alpha\beta,\gamma\delta}_{\mu\nu}(p) = 
  {1\over2} \left(\delta^{\alpha\gamma} \delta^{\beta\delta} - 
                  \delta^{\alpha\delta} \delta^{\beta\gamma}\right)
  (p_\mu p_\nu - p^2 \delta_{\mu\nu}) \widehat{M}(p^2),
\ee
which is not invertible. A propagator for the field 
$A^{\alpha\beta}_\mu$ is obtained by adding a gauge-fixing term, that 
introduces a longitudinal term, makes the matrix invertible, but does not 
contribute to physical quantities \cite{Vasi3}. 

The calculation is completely analogous to that performed in the previous
Section. For the second Legendre transform, we obtain to 
order $O(1/n)$
\be
\Gamma = \Gamma(A=0) + \case{1}{2} {\rm Tr}\, \log D_A + 
   \case{n}{2}\gamma_1(A) + 
   {nm(m-1)\over8} \gamma_2(A) + \ldots
\ee
where $\gamma_1$ and $\gamma_2$ correspond to the graphs reported in 
Fig. \ref{fig-vacuum}, all group-theoretical factors 
have been explicitly singled out, and $\Gamma(A=0)$ is the expression 
reported in Eq. \reff{second-Legendre-transform}.

Generalizing the results of Refs. \cite{Hika,Vasi3} we obtain then
\begin{eqnarray}
&&\rho_1 = \rho_{11}\left[M+{d^2-1 \over 2}(m-1)\right],
\label{rho1-gauge}
\end{eqnarray}
where $M$ is a group-theoretical factor defined in 
Eq. \reff{defM} (for the pure-gauge fixed point $M=0$) and 
$\rho_1$ is the $1/n$ contribution to the exponent $\rho$ defined 
in Eq. \reff{defrho}. Note that the result \reff{rho1-gauge} 
does not depend on the gauge fixing used to define the propagator of the 
field $A$. In principle, one could also compute a gauge-fixing
dependent exponent $\eta$
but its significance is not so clear, since, because of the gauge invariance,
the field $\bphi$ does not have a well-defined anomalous dimension.

As a check we can compare our results with those obtained in 
perturbation theory for the 
gauge Hamiltonian \reff{Hgeneral} with $t_0=0$. Hikami
\cite{Hika} determined the following one-loop 
$\beta$ functions in the $\overline{\rm MS}$ scheme:
\begin{eqnarray}
&&\beta_u =
 -\epsilon u+\left[{m n+8 \over 6}u^2+
   {(m-1)(n-1) \over 6}\left({1 \over 2}v^2-u v\right)-
{3 \over 2}(m-1)u\alpha+{9 \over 8}(m-1)\alpha^2\right],\nonumber \\
&&\beta_v = -\epsilon v+\left[{m+n-8 \over 6}v^2+2 u v
   -{3 \over 2}(m-1)v \alpha+{9 \over 4}(m-2)\alpha^2\right],
\nonumber \\
&&\beta_{\alpha} = -\epsilon\alpha+
   \left[{n \over 12}-{11 \over 3}(m-2)\right]\alpha^2 ,
\end{eqnarray}
where $\alpha = N_d g^2$. These expressions
generalize the results presented in Sec. \ref{sec-epsilon}.
Choosing the $\alpha^\ast=0$ solution of the fixed-point equations we
obtain the four critical points already discussed. However, if we choose the 
solution
\begin{equation}
\alpha^\ast = \epsilon \left[{n \over 12}-{11 \over 3}(m-2)\right]^{-1} + 
    O(\epsilon^2),
\end{equation}
we find another set of four critical points, corresponding to the distinct 
roots of a quartic algebraic equation. This equation cannot be solved in closed 
form, but it is easy to find its roots in the form of a series in the 
powers of $1/n$.
The relevant terms in the expansion of the roots are:
\begin{itemize}
\item chiral-gauge fixed point:
\begin{eqnarray}
&&{u^\ast \over 6 \epsilon} = {1 \over n}+{2-10 m \over n^2} + O(\epsilon),
\nonumber \\
&&{v^\ast \over 6 \epsilon} = {1 \over  n}+{32-10m \over n^2} + 
   O(\epsilon). 
\end{eqnarray}
\item antichiral-gauge fixed point:
\begin{eqnarray}
&&{u^\ast \over 6 \epsilon} = 
   {m-1 \over m n}+{(m-1)(16+88 m-10 m^2) \over m^2 n^2} + O(\epsilon),
\nonumber \\
&&{v^\ast \over 6 \epsilon} = {1 \over  n}+{12+32 m-10m^2 \over m n^2} + 
  O(\epsilon).
\end{eqnarray}
\item Heisenberg-gauge fixed point:
\begin{eqnarray}
&&{u^\ast \over 6\epsilon} = {1 \over m n}+{27 m^3-117m^2+90m-8 \over m^2 n^2}
   + O(\epsilon),
\nonumber \\
&&{v^\ast \over 6 \epsilon} = {27(m-2) \over  n^2} + 
  O(\epsilon).
\end{eqnarray}
\item pure-gauge fixed point:
\begin{eqnarray}
&& {u^\ast \over 6 \epsilon} =  {27(m-1) \over  n^2} + O(\epsilon),
\nonumber \\
&&{v^\ast \over 6 \epsilon} = {27(m-2) \over  n^2} + 
  O(\epsilon).
\end{eqnarray}
\end{itemize}
Thus, the gauge model has in general 8 fixed points and, at least for large
$n$, the chiral-gauge fixed point is the stable one.\footnote{This is not 
true for generic $n$ and $m$. In order to obtain the 
general fixed-point structure,  one should generalize the 
analysis performed in Sec. \ref{sec-epsilon}.}
Substituting these expressions into the relationship \cite{Hika}
\begin{equation}
\nu^{-1} = 2 -{m n+2 \over 6}u^\ast+ 
   {(m-1)(n-1) \over 6}v^\ast+{3(m-1) \over 4}\alpha^\ast + 
   O(\epsilon^2), 
\end{equation}
we find a $1/n$ expanded form of the $O(\epsilon)$ contribution to the critical 
exponent $\nu$ for each of the four solutions:
\be
{1\over\nu} = \cases{
\displaystyle{
2-\epsilon+(48m-42){\epsilon \over n} 
 + O(\epsilon^2,n^{-2})}   &   chiral-gauge f.p.; \cr
 \vphantom{a} & \cr
\displaystyle{
2-{6(m-1)(8m+1) \over m}{\epsilon \over n} + 
     O(\epsilon^2,n^{-2})}   &   antichiral-gauge f.p.; \cr
 \vphantom{a} & \cr
\displaystyle{
2-\epsilon+{45m^2-45m+6 \over m}{\epsilon \over n} 
   + O(\epsilon^2,n^{-2})}   &   Heisenberg-gauge f.p.; \cr
 \vphantom{a} & \cr
\displaystyle{
2-45(m-1){\epsilon \over n} + O(\epsilon^2,n^{-2}).}
        &  pure-gauge f.p.}
\ee
It is then a matter of trivial algebra to verify that these 
expressions are in full agreement 
with the $\epsilon$-expansion of the four solutions 
discussed above in the context of the $1/n$ expansion, which explains 
the names we have given to each fixed point.
Again, we think it is important to notice that the 
conformal bootstrap approach can 
naturally accomodate for the expansion of {\it all} solutions, not only the stable ones.

\section{The $\tilde \epsilon$-expansion of 
$O(n) \times O(m)$ nonlinear $\sigma$ models} \label{sec-sigma}

The LGW Hamiltonian is the natural tool for the study of the critical 
behavior of 
systems near the upper critical dimension $d=4$. If one is interested in 
the critical behavior near the lower critical dimension, one can still 
use perturbation theory, applied however to the 
nonlinear $\sigma$ models (NL$\sigma$). 
The degrees of freedom of the NL$\sigma$ models should 
correspond to the interacting Goldstone modes of the system, 
while the effect of 
the massive modes is only taken into account in the form of constraints for the
massless fields. In this context, it is possible to perform an 
expansion in powers of $\tilde \epsilon\equiv d-2$. 
In the present paper we extend the results of Refs. \cite{Kawa3,Aza0,Aza}
to a general 
$O(n) \times O(m)$ symmetry group and to $O(\tilde \epsilon^2)$. 
Comparing with our previous $1/n$ expansion results we will be able to identify 
the nature of the fixed points of the NL$\sigma$. In particular, we will 
show explicitly that the stable fixed point of the generic model 
can be identified with the stable fixed point of the LGW theory.

We consider the Hamiltonian \reff{Hamiltoniana-sigma}.
This Hamiltonian is geometric in nature, and its variables are best 
understood as generalized coordinates spanning a manifold. The cases we shall 
be interested in correspond to manifolds that are coset spaces. 
More specifically, we must study the coset space (remember that $n\ge m$)
\begin{equation}
{O(n) \times O(m) \over O(n-m) \times O(m)},
\end{equation}
which is topologically equivalent to 
\begin{equation}
{O(n) \over O(n-m)}.
\end{equation}
Associating fields $\pi^I$ with the Goldstone modes that correspond to the broken 
generators 
\begin{equation}
\{ {\rm Lie}\ (O(n))-{\rm Lie}\ (O(n-m))\},
\end{equation}
the Hamiltonian may be formulated in purely geometric terms, i.e.
\begin{equation}
\tilde H = {1 \over 2} g_{IJ}(\pi) \nabla \pi^I \nabla \pi^J.
\end{equation}
The couplings $T_i \equiv 1/\eta_i$ are 
related to the independent entries 
of the tangent-space metric $\eta_{IJ}$.

A number of important RG properties of the NL$\sigma$ 
models have been derived in the general case by Friedan \cite{Fried} 
and specialized to the models of 
interest in Refs.  \cite{Aza0,Aza}. 
If $R_{IJKL}$ and $R_{IJ}$ are respectively the 
Riemann and Ricci tensor for the metric $g_{IJ}$,
the RG $\beta$ functions of the model can be written to two-loop 
order as
\begin{equation}
\beta_{IJ} \equiv s {\partial g_{IJ} \over \partial s} = - \tilde \epsilon \eta_{IJ} + 
R_{IJ}+ {1 \over 2}R_{IPQR}R_J^{\;PQR} + \dots
\end{equation}
The number
of algebraically independent  $\beta$-functions $\beta_i$ coincides with the
number of independent couplings $T_i$.  Therefore we should consider 
two $\beta$-functions associated with $T_1$ and $T_2$. The fixed points are 
determined from the equations
\begin{equation}
\beta_i(T_1^\ast,T_2^\ast)=0,
\end{equation}
that can be perturbatively solved in powers of $\tilde \epsilon$.

The evaluation of the two-loop $\beta$ functions for 
arbitrary $m$ and $n$ requires no special 
skills, but it takes some time and effort in view of the many 
computational steps involved.
Without belaboring on the intermediate steps, we report here our 
final results:
\begin{eqnarray}
\hskip -0.5truecm
&&\beta_1 \equiv -s{\partial T_1 \over \partial s}= -\tilde \epsilon
 T_1+\left[n-2-{m-1 \over 2}X\right]T_1^2 +
  \left[A+B X+C X^2\right]T_1^3+ O(T_i^4),
\label{beta-sigma} \\
\hskip -0.5truecm
&&\beta_2 \equiv -s{\partial T_2 \over \partial s}= 
-\tilde \epsilon T_2+\left[{m-2 \over 2} +
{n-m \over 2}X^2\right]T_2^2 +\left[D X^2+E X^3+F X^4+G\right]T_2^3+ 
   O(T_i^4), \nonumber
\end{eqnarray}
where $X$ is shorthand for the ratio $T_1/T_2$ and we have defined the coefficients:
\begin{eqnarray}
&&A(n,m) \equiv 2m(n-m)-n+{3 \over 8}(m-1)(m-2), \\
&&B(n,m) \equiv -(m-1)\left[{3 \over 2}(n-m)+{3 \over 8}(m-2)\right],\nonumber \\
&&C(n,m) \equiv (m-1)\left[{3 \over 8}(n-m)+{m \over 8}\right],\nonumber \\
&&D(n,m) \equiv {3 \over 4}(n-m)(m-2),\nonumber \\
&&E(n,m) \equiv -{3 \over 4}(n-m)(m-2),\nonumber \\
&&F(n,m) \equiv {1 \over 8}(n-m)(2m-3),\nonumber \\
&&G(n,m) \equiv {1 \over 8}(m-2)^2.\nonumber
\end{eqnarray}
Our results were submitted to a number of basic consistency checks:

i) when $m=1$ there is no $\eta_2$ coupling, and $\beta_1(T_1)$ 
reduces to the well-known $\beta$ function for the 
vector $\sigma$-model defined on 
the coset space $O(n)/O(n-1)$.

ii) when $m=2$ our expressions reduce to those of Ref. \cite{Aza}.

iii) when $m=n$ there is no $\eta_1$ coupling,\footnote{If $n=m$ 
the vectors ${\bf e}^\alpha$ are an orthogonal basis in $R^m$ and therefore
satisfy the completeness relation $\sum_\alpha ({\bf v}\cdot {\bf e}^\alpha)
({\bf w}\cdot {\bf e}^\alpha) = {\bf v}\cdot {\bf w}$ for all vectors 
${\bf v}$, ${\bf w}$. Then, it is a simple matter to show that 
the second term in Eq. \reff{Hamiltoniana-sigma} is one half of the 
first one.}
and the model reduces to a standard 
$O(n) \times O(n)$ principal chiral model.
One may verify that $\beta_2(T_2)$ is 
directly related to the known $\beta$ function of these models 
\cite{Hikami-81}.

One may also consider the ``gauge'' limit $\eta_2=0$, which was studied 
by Hikami \cite{Hika}: 
the identification with Hikami's coupling is $\eta_1 \rightarrow 1/t$.
One must however recognize that the limit is singular, and as a
 consequence the function $\beta_1^{gauge}(t)$ is {\it not} obtainable 
from our expressions by setting $X=0$ 
(with the notable exception of $m=2$ models).
If we assume $\eta_2=0$ from the very beginning of our calculation, 
the result is
\begin{equation}
\beta_1^{gauge}(t) = -\tilde \epsilon t+ (n-2)t^2+[2m(n-m)-n]t^3+O(t^4),
\end{equation}
consistent with that reported in Ref. \cite{Hika}.

The $\beta$-functions \reff{beta-sigma} are the starting point for 
the perturbative 
evaluation of the critical points and exponents to $O(\tilde \epsilon^2)$.
A consistent Ansatz for the simultaneous solutions of the equations 
$\beta_i(T_1^\ast,T_2^\ast)=0$ is the following:
\begin{eqnarray}
&&T_1^\ast= t_1 \tilde \epsilon+ t_2 \tilde \epsilon^2+ O(\tilde \epsilon^3),
\\
&&X^\ast= X_0+X_1 \tilde \epsilon+ O(\tilde \epsilon^2).
\end{eqnarray}
It is straightforward to obtain the following algebraic equations for $t_1$ and $X_0$:
\begin{eqnarray}
&&-{1 \over t_1}+ (n-2)-{m-1 \over 2}X_0=0,\\
&&-{1 \over t_1}X_0 +{m-2 \over 2}+ {n-m \over 2}X_0^2 =0.
\end{eqnarray}
They are trivially solved by
\begin{eqnarray}
&&X_0^{\pm}={n-2 \pm \sqrt{(n-2)^2-(n-1)(m-2)} \over n-1}, \nonumber \\
&&{1 \over t_1^{\pm}}= n-2-{m-1 \over 2}X_0^{\pm} .
\label{fixed-points-sigma}
\end{eqnarray}
Iterating the procedure we may also obtain
\begin{eqnarray}
&&X_1^\pm= {(D-A)(X_0^\pm)^2+(E-B)((X_0^\pm)^3+(F-C)(X_0^\pm)^4+G \over X_0^\pm 
[(n-2)-(n-1)X_0^\pm]\left(n-2-{m-1 \over 2}X_0^\pm\right)},\\
&&t_2^\pm = \left[{m-1 \over 2}X_1^\pm-{A +B X_0^\pm+ C(X_0^\pm)^2 \over n-2-
{m-1 \over 2}X_0^\pm}\right]{1 \over \left(n-2-{m-1 \over 2}X_0^\pm\right)^2}.
\end{eqnarray}
This analysis shows the existence of a couple of nontrivial 
fixed points of the RG equations. 
However, it is evident from Eq. \reff{fixed-points-sigma} that such a pair 
of solutions does not exist for all $m$ and $n$: for some values $X_0^\pm$ 
is indeed complex. Repeating the analysis we performed in Sec. 
\ref{sec-epsilon} for the $\epsilon$-expansion, we see that these two fixed 
points exist only 
for $n > \widetilde{n}^+$ and $n< \widetilde{n}^-$, where
\begin{equation}
\widetilde{n}^\pm = 
  {m+2\pm\sqrt{m^2-4} \over 2}\pm{1 \over 2}\sqrt{{m+2 \over m-2}}
   {m^2+4\pm m\sqrt{m^2-4}
 \over (1\pm \sqrt{m^2-4})(m\pm\sqrt{m^2-4})}\tilde \epsilon+ 
O(\tilde \epsilon^2).
\label{ntildepm}
\end{equation}
Note that for $\tilde \epsilon$ small, we have 
$m\le \widetilde{n}^+ < m+ 1$ and $\widetilde{n}^- < m$. 
Thus, since $n\ge m$, all models with integer $n \ge m +1$ have a 
a pair of nontrivial fixed points, at least for $\tilde\epsilon$ small. 
Beside these two fixed points, there is also a fixed point 
for $T_1=0$ and $T_2= T_2^*$ that belongs to the universality class 
of the $O(m)\times O(m)$ principal chiral model. Such a fixed point 
always exists perturbatively for $m>2$ and in particular is the 
only resent for $n=m$.

As one may easily notice, the expansion \reff{ntildepm} is singular when $m=2$.
This is related to the 
following peculiar feature of $m=2$ models:
for any value of $n$ the (unstable) fixed point 
corresponds to the solution $X^\ast =0$, 
and as a consequence we observe 
its coalescence with the ``gauge'' fixed point obtained 
by setting $\eta_2 =0$. This phenomenon does not happen for $m>2$. 
In this case, the gauge fixed point and the antichiral fixed point are
distinct.

The exponent $\eta$ is easily computed. To two-loop order we have
\begin{equation}
\eta =  -\tilde \epsilon+(n-m)T_1^\ast + (m-1)T_2^\ast+O(\tilde \epsilon^3).
\end{equation}
Substituting the expression of the fixed point, we obtain 
\begin{equation}
\eta = (n-m)(t_1 \tilde \epsilon+ t_2 \tilde \epsilon^2)+
   (m-1)\left[{t_1 \over X_0}
\tilde \epsilon+
   \left({t_2 \over X_0}-{t_1 X_1 \over X_0^2}\right)\tilde \epsilon^2\right]-
\tilde\epsilon+O(\tilde \epsilon^3).
\end{equation}
We can expand the $\eta$ 
at the stable critical point in powers of $1/n$, obtaining  
\begin{equation}
\eta =  \left[{m+1 \over 2n}+ {3m^2+7m+6 \over 8n^2}+
 O\left({1 \over n^3}\right)\right]\tilde \epsilon
   - \left[{m+1 \over 2n}+{3(m+1)^2 \over 4n^2}
+O\left({1 \over n^3}\right)\right]\tilde \epsilon^2 +O(\tilde \epsilon^3).
\end{equation}
If we compare such expression
with the $\tilde \epsilon$-expansion of $\eta^+$ as obtained from the 
large-$n$ expansion 
of Sec. \ref{sec-largen}, Eq. \reff{etapiu-largen}, we find complete agreement, 
confirming the identification of the two fixed points.

In NL$\sigma$ models the evaluation of stability goes together 
with the evaluation of the critical exponent $\nu$, since both are 
related to the eigenvalues of the 
derivative matrix 
\begin{equation}
{\partial \beta_i \over \partial T_j}(T_1^\ast,T_2^\ast).
\end{equation}
More precisely, stability requires that the above matrix possesses only 
one positive eigenvalue $\lambda_+ = \nu^{-1}$.
The presence of two positive eigenvalues signals the instability of the 
fixed point.  It is possible to evaluate the above-mentioned eigenvalues 
in the context of the $\tilde \epsilon$ expansion, obtaining:
\begin{eqnarray}
&&\lambda_+ = \tilde \epsilon-\nu_2(n,m)\tilde \epsilon^2+
O(\tilde \epsilon^3),\\
&&\lambda_- = \lambda_1(n,m)\tilde \epsilon+\lambda_2(n,m)\tilde \epsilon^2+
O( \tilde \epsilon^3),
\end{eqnarray}
where
\begin{eqnarray}
&&\nu_2(n,m)= {a_{112}a_{221}+a_{111}a_{222}-a_{122}a_{211}-a_{121}a_{212} 
\over a_{111}+a_{221}}, \nonumber \\
&&\lambda_1(n,m)= 1-a_{111}-a_{221},\nonumber\\
&&\lambda_2(n,m)=\nu_2(n,m) -a_{112}-a_{222}.
\end{eqnarray}
Here we defined
\begin{eqnarray}
&&a_{111} \equiv {m-1 \over 2}t_1 X_0 = -a_{121}X_0,\\
&&a_{211} \equiv  -(n-m)t_1 X_0^2= -a_{221}X_0,\nonumber\\
&&a_{112} \equiv 
   {m-1 \over 2}(t_1 X_1+t_2 X_0)-t_1^2(A+2B X_0+3C X_0^2),\nonumber\\
&&a_{122} \equiv  -{m-1 \over 2}t_2+ t_1^2(B+2C X_0),\nonumber\\
&&a_{212} \equiv  
-(n-m)(2 t_1 X_0 X_1+t_2 X_0^2)-t_1^2(2D X_0+3E X_0^2+4 F X_0^3),\nonumber\\
&&a_{222} \equiv  
(n-m)(t_1 X_1+ t_2 X_0)+ t_1^2(D +2E X_0+3 F X_0^2-G/X_0^2).\nonumber
\end{eqnarray}
The $1/n$ expansion of $\nu$ evaluated at the stable
fixed point coincides with the $\tilde \epsilon$ 
expansion of $\nu^+$ obtained in Sec. \ref{sec-largen}.
The result of the expansion is:
\begin{equation}
\nu_2 = {m+1 \over 2}{1 \over n}+ {(m+1)^2 \over 2} {1 \over n^2} + 
  O\left({1 \over n^3}\right).
\end{equation}
Notice that the coalescence value $\tilde{n}^\pm$ can be easily determined
within the 
$\tilde \epsilon$ expansion by imposing the condition
\begin{equation}
\lambda_-[\tilde{n}^\pm,m]= 0.
\end{equation}
While the stable fixed point is identified with the chiral fixed point of the 
$O(n)\times O(m)$ LGW model, the unstable one is unrelated to those
of the LGW model. In order to understand its nature, it is again useful 
to consider the large-$n$ limit. From Eq. \reff{fixed-points-sigma} 
one observes that $X_0^- \sim 1/n$ as $n\to \infty$. Therefore, the fixed 
point survives in the large-$n$ limit if we scale the coupling constants as 
$T_1 = O(1/n)$, $T_2 = O(1)$. Then, for large $n$
the $\beta$ functions decouple. Moreover, while $\beta_1(T_1)$ 
gets no contributions beyond 
one loop (as usual in vector models), 
$\beta_2(T_2)$ turns into the $\beta$ function of an 
$O(m) \times O(m)$ principal chiral model \cite{Hikami-81}. Therefore,
in this case the pattern of spontaneous 
symmetry breaking is highly nontrivial,
even in the strict $n \rightarrow \infty$ limit. 
This can be understood from Eqs. \reff{H-largen} and \reff{Heff}. In the 
large-$n$ limit the relevant Hamiltonian is 
\be
H = H_{\rm eff} + {t_0\over2} A^2.
\ee
Now, the field $\bphi$ couples only to the gauge-invariant degrees of 
freedom, and thus at the saddle point the field $A_\mu^{\alpha\beta}$ 
is a pure gauge transformation, i.e. 
\be
A_\mu^{\alpha\beta} = \left( O^{-1} \partial_\mu O \right)^{\alpha\beta},
\ee
where $O$ is an $O(m)$ matrix. Thus, for $n\to \infty$, the Hamiltonian
can be rewritten as the sum of two terms: 
\be 
H = {1\over2} \sum_\alpha 
  \bphi_\alpha \cdot (\partial_\mu \partial_\mu + M^2) \bphi_\alpha + 
  {t_0\over 2} {\rm Tr}\ \partial_\mu O^{-1} \partial_\mu O,
\ee
where, as in Sec. \ref{sec-largen}, $M^2 = r_0 + w_0 \sigma$.
Thus, the unstable fixed point is directly related to the nontrivial 
fixed point of the principal $O(m)\times O(m)$ chiral model. 

Finally, we consider the gauge limit
$\eta_1 = 1/t$, $\eta_2 = 0$. We find a nontrivial fixed point:
\begin{equation}
t^\ast = 
  {1 \over n-2}\tilde \epsilon+{n-2 m(n-m) \over (n-2)^3}\tilde \epsilon^2+ 
O(\tilde\epsilon^3).
\end{equation}
Correspondingly we obtain:
\begin{equation}
\nu^{-1} \equiv \beta^\prime(t^\ast) = 
\tilde\epsilon+ {2 m(n-m)-n \over (n-2)^2}\tilde\epsilon^2+
 O(\tilde\epsilon^3).
\end{equation}
The $1/n$ expansion of the above result gives
\begin{equation}
\nu^{-1} =  \tilde\epsilon+ {2m-1 \over n}\tilde\epsilon^2 +
 O(\tilde\epsilon^3),
\end{equation}
and one may easily check that it agrees with the $\tilde\epsilon$ expansion 
of the results found in Sec. \ref{sec-vector}
for the stable fixed point of the gauge model, see Eq. \reff{rho1-gauge},
with $M = M^+$.

The behavior of NL$\sigma$ models in the case $m=n-1$ is worth a 
special discussion \cite{DJ-96}. 
In this case one may naturally define  a new $n$-component field  
${\bf e}_n$ such that 
${\bf e}_n  \cdot {\bf e}_n = 1$ and  
${\bf e}_\alpha  \cdot {\bf e}_n = 0$, and one can show that
\begin{equation}
\tilde H = {1 \over 4}\eta_2 
   \sum_{\alpha=1}^{n-1} 
   \partial_\mu{\bf e}_\alpha \cdot \partial_\mu{\bf e}_\alpha + 
\left({1 \over 2}\eta_1-{1 \over 4}\eta_2\right) 
\partial_\mu{\bf e}_n \cdot \partial_\mu{\bf e}_n.
\end{equation}
When $\eta_1=\eta_2$,
this is the Hamiltonian of an $O(n) \times O(n)$ principal chiral model.
The stable fixed point is characterized by the property that $X^\ast=1$ and one finds:
\begin{equation}
t_1={2 \over n-2}, \qquad t_2=-{1 \over n-2}.
\end{equation}
Direct substitution shows that
\begin{eqnarray}
&&\nu^{-1} = \tilde \epsilon+ {1 \over 2}\tilde \epsilon^2+ O(\tilde \epsilon^3),\\
&&\eta ={n \over n-2}\tilde \epsilon-{n-1 \over n-2}\tilde \epsilon^2 +O(\tilde \epsilon^3).
\end{eqnarray}
As one may easily check, these exponents coincide with those obtained in the case $m=n$.
Therefore, as shown in Ref. \cite{DJ-96},
the symmetry of the  $O(n) \times O(n-1)$ model is dynamically promoted to  
$O(n) \times O(n)$ at the stable fixed point.
This property is certainly true for sufficiently small $d > 2$, but at this level of analysis 
it is impossible to establish the maximum dimension $\hat{d}_c$ for which a 
stable critical point 
possessing the enlarged symmetry can be found.

\section{Conclusions} \label{sec-conclusions}

At this stage of our analysis,
we can draw quite general conclusions on the general fixed-point 
structure of the models with $O(n) \times O(m)$ 
for all dimensions $2 \leq d \leq 4$.
By comparing the $\tilde{\epsilon}$-expansion results near two
dimensions, the $\epsilon$-expansion results near four dimensions 
and the large-$n$ results, we have been able to identify the nature of 
all (stable and unstable) fixed points of these models. In particular, 
the LGW stable fixed point coincides with the stable one of 
the NL$\sigma$ model. We thus quantitatively confirm one of the conclusions of
Ref. \cite{Aza0,Aza}: above
$\bar n(m,d)$ in the $(m,d)$ plane a second-order phase transition occurs,
which is described by the common fixed point of the 
NL$\sigma$ and LGW models. 

The unstable fixed points, which give rise to 
different types of tricritical behavior and crossover phenomena,
are instead unrelated and
correspond to systems with completely different types of excitations.

The correspondence we have found holds only for sufficiently large values of 
$n$, i.e. for $n \ge \bar{n}(m,d)$, which is the region of analyticity of the 
large-$n$ expansion. 
Since the $1/n$ expansion commutes with the  $\epsilon=4-d$ expansion of the
LGW Hamiltonian, 
one may expect $\bar n(m,d)$ to coincide with $n^+(m,d)$ in a neighborhood of 
$d=4$; $n^+(m,d)$ might in turn be evaluated within the $1/n$ expansion by 
solving the coalescence equation:
\begin{equation}
\nu^+(m, n^+,d)= \nu^-(m,n^+ ,d).
\label{coalescenza-largen}
\end{equation}
The estimate obtained from the lowest order approximation for $\nu^\pm$ 
is:\footnote{This expression has been obtained by solving exactly the equation
$1/\nu^+(m, n^+,d) = 1/\nu^-(m,n^+ ,d)$, where $1/\nu^+(m, n^+,d)$
and $1/\nu^-(m,n^+ ,d)$ are expanded to order to $1/n$. }
\begin{equation}
n^+(m,d) \approx 2\left(m+1- {1 \over m}\right){\rho_{11}(d) \over 4-d}.
\end{equation}
This expression shows the correct qualitative behavior for all 
$2 \leq d \leq 4$  and a rough quantitative agreement. 
It is possible to improve the approximation
by including the $1/n^2$ correction in Eq. \reff{coalescenza-largen}. 
For $m=2$ and $d=3$, it predicts 
$n^+(2,3) \approx 5.3$, in substantial  agreement with the results obtained 
by using the ERG approach  \cite{Tiss2,Tiss3}, the perturbative expansion
in fixed dimension \cite{PRV-00}, and, as we shall show below, the 
$\epsilon$-expansion.

\begin{figure}[tb]
\centerline{\psfig{width=6truecm,angle=0,file=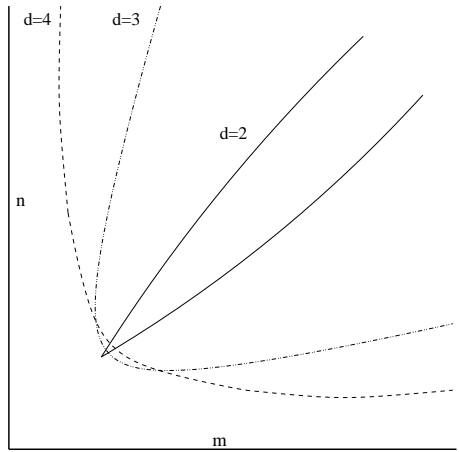}}
\vspace{0cm}
\caption{Sketch of the large-$n$ analyticity boundary 
as a function of the dimension $d$. }
\label{fig2}
\end{figure} 

We want now to understand the behavior of $\bar{n}(m,d)$ near two dimensions.
Near two dimensions, using the NL$\sigma$ model results we know that 
the (LGW and  NL$\sigma$) stable fixed point exists only 
for $n > \widetilde{n}^+ (m,d)$, so that in this case 
$\bar{n}(m,d) = \widetilde{n}^+ (m,d)$. Thus, for generic values of $d$ 
we conjecture
\be
\bar{n}(m,d)  = \cases{n^+(m,d) & for $d_c(m) \le d \le 4$\cr
                       \widetilde{n}^+ (m,d) &  for $2 \le d \le d_c(m)$,}
\ee
where $d_c(m)$ is a critical dimension that we cannot determine with our means.
Of course, this expression is valid for $n\ge m$. The symmetry 
under exchange of $n$ and $m$, implies the existence 
of a similar boundary curve in the region $n \le m$, obtained by interchanging 
$n$ and $m$.
A sketch of $\bar{n}(m,d)$ is reported in Fig. \ref{fig2}.

Now, let us discuss the behavior of the LGW fixed points for $d \to 2$.
Since the LGW stable fixed point
is equivalent to that of the NL$\sigma$ model, and, for all $n\ge 2$, 
$m\ge 2$ except $n=2$, $m=2$, the NL$\sigma$ model is asymptotically
free, we expect  $\nu^+ = \infty$, a 
conclusion that is confirmed by the large-$n$ expression 
\reff{nupiu-largen}. On the other hand, for $d=2$, Eq. \reff{numeno-largen}
predicts $\nu^- = 1/2$ without $1/n$ and $1/n^2$ corrections. 
It is thus natural to conjecture that $\nu^- = 1/2$ for 
all $n\ge2$ and $m\ge2$,
i.e. that the LGW antichiral fixed point is a Gaussian fixed point. 
The case $m=2$, $n=2$ needs a special discussion. Using the fact that 
the $O(2)\times O(2)$ LGW model is 
equivalent to the so-called $mn$ model 
\footnote{
The $mn$ model with $m=2$ describes $n$ $XY$ models coupled by an
O($n$)-symmetric interaction. 
Using essentially
nonperturbative arguments (see e.g. Ref. \cite{Aharony-76})
related to the specific-heat exponent of the 
$XY$ universality class, one can argue 
that for $d<d_c$ with $d_c>3$ there is a stable fixed point belonging
to the $XY$ universality class,
while for $d>d_c$ there is a stable fixed point with the tetragonal symmetry.
This fact has been recently confirmed by high-order
field-theoretical calculations in three dimensions \cite{PV-00-r,PV-01}.}
with $m=n=2$ \cite{AS-94,Aharony-76}, 
one can show that 
in the $v<0$ region
a stable fixed point exists for all values of $d$ \cite{KawaR,PV-01}. 
Finally, for $d\to 2$, using the $\sigma$-model
results of the previous Section, one finds that it becomes Gaussian. 

We  want now to use the knowledge of $\bar{n}(m,2)$ in order to 
obtain some informations on $\bar{n}(m,3)$. For this purpose we will
make two hypotheses: first we will assume $d_c(m) < 3$, so that 
$\bar{n}(m,3) = n^+(m,3)$; second, we will assume $\bar{n}(m,d)$ 
to be sufficiently smooth in $d$ at $m$ fixed, so that we can
use the interpolation method of Ref. ~\cite{LGZ-87}.
Such a method has provided very precise estimates of critical 
quantities (see, e.g., Refs.~\cite{LGZ-87,PV-98}). 

Let us first consider the case $m=2$. 
We start from \cite{ASV-95}
 \begin{eqnarray}
\bar{n}(2,4-\epsilon) &=  & 12 +4 \sqrt{6}-
      \left(12+{14 \over 3} \sqrt{6}\right)\epsilon
\nonumber \\
&& +\left[{137 \over 150}+ {91 \over 300}\sqrt{6}+
     \left({13 \over 5}+{47 \over 60}\sqrt{6}\right)
   \zeta (3)\right]\epsilon^2  + O(\epsilon^3)
\nonumber \\[2mm]
&= & 21.80 -23.43 \epsilon + 7.09 \epsilon^2 + O(\epsilon^3).
\end{eqnarray}
Following Ref.~\cite{LGZ-87},
we rewrite this equation in the following form 
\be
\bar{n}(2,4-\epsilon) = 2 + (2- \epsilon) 
  \left(9.90 - 6.67 \epsilon
     + 0.16 \epsilon^2 \right) + O(\epsilon^3).
\label{npiu-constrained}
\ee
Note that 
the new perturbative series is much better behaved than the original one, 
the coefficients of the series decreasing rapidly. 
Setting $\epsilon = 1$, we obtain an estimate 
for $\bar{n}(2,3)$:
\be 
\bar{n}(2,3) \approx  5.3(2),
\label{npiu23}
\ee
where the ``error" indicates how the estimate varies from two loops to 
three loops. It should not be taken seriously; it should only provide 
an order of magnitude for the precision of the results. 
The estimate \reff{npiu23} is in  good agreeement
with the determinations of Ref. \cite{Tiss3,PRV-00}: 
$\bar{n}(2,3) \approx 5$ (Ref. \cite{Tiss3}), 
$\approx 6$ (Ref. \cite{PRV-00}). 
We can try to estimate the exponents for $n=\bar{n}(2,3)$, by using 
Eq. \reff{exponents-ncritico}. The coefficients decrease steadily 
with $\epsilon$ and thus we can simply set $\epsilon=1$, obtaining
\be
\eta\approx 0.038, \qquad\qquad \nu \approx 0.63.
\ee
It is also interesting to compute the exponents for $n = 6$, $m=2$, and $d=3$,
in order to make a numerical comparison with the results of 
Ref. \cite{LSDASD-00} who found $\nu = 0.700(11)$, 
$\gamma = 1.383(36)$, and the ERG results of Ref. \cite{Tiss3} 
who found $\nu \approx 0.707$, $\gamma \approx 1.377$. If we use our
$O(n^{-2})$ expansions for the critical exponents, we obtain 
$\gamma \approx 1.22$ and $\nu \approx 0.63$.
We can also use the 
$\epsilon$-expansion, by using the method explained at the end of 
Sec. \ref{sec-epsilon}. In this case, we must fix 
$\Delta n = 6 - \bar{n}(2,3)$. Conservatively, we have
$0 \ltapprox \Delta n \ltapprox 1$. Then, from the perturbative series
we estimate $\nu\approx 0.63$ - 0.64, $\gamma \approx 1.24$ - 1.26,
which is rather close to the large-$n$ result, and   
somewhat lower than the numerical results of
Refs.~\cite{LSDASD-00,Tiss3}.  
It is also worth mentioning that for $n=6$ the
fixed-dimension field-theoretical approach
do not find fixed points that are sufficiently stable with respect to the 
order of the expansion up to six loops \cite{PRV-00}. 
We believe that these apparent discrepancies among the various
approaches  deserve further investigation.

Our expressions may also be employed in order to compute the 
critical dimensionality $\hat{d}_c$ such that, for $d<\hat{d}_c$
the $O(3)\times O(2)$ has a nontrivial fixed point with symmetry 
$O(4)$ \cite{Aza0,Aza}. This corresponds to solving the equation
$\bar{n}(2,\hat{d}_c)=3$. If we use Eq. \reff{npiu-constrained}, we  find
$d_c \approx 2.71$. 
We may compare our result to those obtained in the ERG approach:
$\hat{d}_c =$ 2.83 (Ref. \cite{Tiss1}), 2.87 (Ref. \cite{Tiss2}).
They are in substantial agreement with our result, when allowing
for the systematic errors of both approaches. 
It should also be noticed that our interpolation \reff{npiu-constrained} 
is also in very good agreement with the ERG results of Refs. \cite{Tiss1,Tiss2}
for all $\epsilon$ \cite{Delamotte-private}.

These analyses can be repeated for larger values of $m$. Since only 
$\bar{n}(m,3)$ seems to be rather precisely determined, we only report the 
results for this quantity. For $m=3$ and $4$ we have the 
constrained estimates
\begin{eqnarray} 
   \bar{n}(3,3) &\approx & 9.1(9), \\
   \bar{n}(4,3) &\approx & 12(1).  
\end{eqnarray}
For large $m$ we have
\begin{eqnarray}
\bar{n}(m,4-\epsilon) &=& m \left(
   9.90 - 10.10 \epsilon + 2.66 \epsilon^2 + O(\epsilon^3,m^{-1})\right) 
\nonumber \\
 &=& m + (2-\epsilon) m \left[
   4.45 - 2.83 \epsilon
  - 0.084 \epsilon^2  + O(\epsilon^3,m^{-1})\right],
\end{eqnarray}
where, as already observed, the coefficients of the constrained series 
are smaller than the original ones. Setting $\epsilon = 1$, we obtain 
$\bar{n}(m,3) \approx 2.5 m $. Note that the large-$m$ approximation is 
already good at $m=4$.

It is very important to notice that, 
since all extrapolation techniques are adiabatic 
in their parameters, 
it is not possible to catch the (essentially nonperturbative) features of 
the models in the region below $\bar n$.
As a consequence there is no inconsistency between 
the present statements and our results  
\cite{PRV-00} concerning $O(n) \times O(2)$ models for $n=2$, 3 in fixed 
dimension $d=3$.  The fixed points we found 
for $n=2,3$ are certainly not analytically connected with the 
large-$n$ and small-$\epsilon$ criticalities discussed in
this paper.


Finally, we recall that an enlarged parameter space for the 
$O(n) \times O(m)$ symmetric models
 with critical dimension $d=4$ leads to the 
appearance of several new, generally unstable, fixed points, that 
physically correspond to tricritical transitions and give rise to 
crossover phenomena.
It is important to recognize that the conformal bootstrap approach to the $1/n$ expansion allows 
a consistent treatment of all these criticalities.
Systems with $O(n) \times O(m)$ symmetry may also possess a ``gauge'' criticality, which can
 be described by the appropriate $1/ n$ expansion as well as 
within the $\epsilon$ expansion of the Hamiltonian for 
scalar chromodynamics and within 
the $\tilde\epsilon$ expansion of a class 
of gauge-invariant NL$\sigma$ models.

\bigskip 

\bigskip 

{\bf Acknowledgements}

\bigskip
We thank B. Delamotte, D. Mouhanna, and M. Tissier for useful correspondence.

\end{document}